\documentclass{aa}
\usepackage{graphicx}
\usepackage{txfonts}
\usepackage{natbib}
\usepackage{color}
\bibpunct{(}{)}{;}{a}{}{,}    
\newcommand{\beq}{\begin{equation}}
\newcommand{\eeq}{\end{equation}}
\newcommand{\bea}{\begin{eqnarray}}
\newcommand{\eea}{\end{eqnarray}}
\newcommand{\lsim}{\raisebox{-0.6ex}{$\stackrel{{\displaystyle<}}{\sim}$}}

\newcommand{\url}[1]{{\tt #1}}

\def\gapp{\lower 3pt\hbox{${\buildrel > \over \sim}$}\ }
\def\lapp{\lower 3pt\hbox{${\buildrel < \over \sim}$}\ }

\usepackage{color}
\usepackage{ulem}

\newlength{\linwx}
\begin{document}
\title{Influence of viscosity and the adiabatic index on planetary migration}
\author{
Bertram Bitsch \inst{1,2},
Aaron Boley \inst{3}
\and
Wilhelm Kley  \inst{1}
}
\offprints{B. Bitsch,\\ \email{bertram.bitsch@oca.eu}}
\institute{
     Institut f\"ur Astronomie \& Astrophysik, 
     Universit\"at T\"ubingen,
     Auf der Morgenstelle 10, D-72076 T\"ubingen, Germany
\and
Dep. Cassiop\'{e}e, University of Nice-Sophia Antipolis, CNRS, Observatoire de la C\^{o}te d'Azur, 06304, Nice
\and
University of Florida, Department of Astronomy, 211 Bryant Space Science Center, Gainesville, FL 32611, USA
}
\abstract
{ The strength and direction of migration of low mass embedded planets depends on the disk's thermodynamic state. 
It has been shown that in active disks, where the internal dissipation is balanced by radiative transport, migration can be directed outwards, a process which extends the lifetime of growing embryos. 
Very important parameters determining the structure of disks, and hence the direction of migration, are the viscosity and the adiabatic index. 
 } 
{ In this paper we investigate the influence of different viscosity prescriptions ($\alpha$-type and constant) and adiabatic indices on disk structures. 
We then determine how this affects the migration rate of planets embedded in such disks.
} 
{ We perform three-dimensional numerical simulations of accretion disks with embedded planets. 
We use the explicit/implicit hydrodynamical code {{\tt NIRVANA}} that includes full tensor viscosity and radiation transport in the flux-limited diffusion approximation, as well as a proper equation of state for molecular hydrogen. 
The migration of embedded 20 $M_{\rm Earth}$ planets is studied.
 } 
{
 Low-viscosity disks have cooler temperatures and the migration rates of embedded planets tend toward the isothermal limit.
  Hence, in these disks, planets migrate inwards even in the fully radiative case. 
  The effect of outward migration can only be sustained if the viscosity in the disk is large. 
  Overall, the differences between the treatments for the equation of state seem to play a more important role in disks with higher viscosity. A change in the adiabatic index and in the viscosity changes the zero-torque radius that separates inward from outward migration.
}
{
 For larger viscosities, temperatures in the disk become higher and the zero-torque radius moves to larger radii, allowing outward migration of a 20-$M_{\rm Earth}$ planet to persist over an extended radial range. 
 In combination with large disk masses,  this may allow for an extended period of the outward migration of growing protoplanetary cores.
}
\keywords{accretion disks -- planet formation -- hydrodynamics -- radiative transport -- planet disk interactions -- adiabatic index}
\maketitle
\markboth
{Bitsch, Boley \& Kley: Influence of viscosity and the adiabatic index on planetary migration}
{Bitsch, Boley \& Kley: Influence of viscosity and the adiabatic index on planetary migration}

\section{Introduction}
\label{sec:introduction}

Migration of a low-mass planet embedded in a fully radiative gaseous disk can be significantly different from migration in an isothermal or purely adiabatic disk \citep{2006A&A...459L..17P, 2008ApJ...672.1054B, 2008A&A...485..877P, 2008A&A...478..245P, 2008A&A...487L...9K, 2009A&A...506..971K, 2010MNRAS.408..876A}. 
While all authors agree that radiation transport can slow the rate of inward migration, there is still a lack of consensus whether the direction of migration can be outward. 
Part of this confusion may be due to the sensitivity of the direction and magnitude of migration on global disk parameters \citep{2010MNRAS.401.1950P, 2010ApJ...723.1393M, 2011MNRAS.410..293P}, including, e.g., the radial disk temperature gradient \citep{2011arXiv1103.3502A}. 
Different authors also use different viscosities, usually either a constant viscosity or a \citet{1973A&A....24..337S} $\alpha$-viscosity, with a typical value of $\alpha$ ranging between $10^{-4}$ and $10^{-2}$. 
The simulation of an embedded planet in a global magnetorotationally unstable accretion disk through direct magnetohydrodynamical calculations including radiative transport is still computationally too expensive. 
However, the disk structure of magnetorotationally unstable disks including radiative transport has been calculated in the local shearing box approximation \citep{2012MNRAS.tmp.2186F}.

An unperturbed, viscous, fully radiative disk will evolve towards an equilibrium state, where viscous heating is balanced by radiation transport and cooling \citep[as described in, e.g., ][]{2009A&A...506..971K}. 
This equilibrium state is dependent on the disk mass, the viscosity, and the adiabatic index of the gas. 
Variations in viscosity change the radial density and temperature profiles, thus changing the profiles of vortensity and entropy. 
The vortensity and entropy gradients across the horseshoe region, in turn, determine the magnitude and sign of the corotation torque \citep{2008ApJ...672.1054B}. 
If the torques can remain unsaturated due to the action of viscosity, the total torque can be positive, leading to outward migration. 
This effect is also possible in isothermal disks \citep{2008A&A...485..877P}.  

There is an additional complication to understanding the migration of low-mass planets in disks. 
The rotational states of molecular hydrogen are only fully accessible at temperatures $\gtrsim 300$ K \citep{1978ApJ...223..854D, 2007ApJ...656L..89B}. 
As a result, the adiabatic index of the gas will transition from $\gamma=7/5$ to $5/3$ as the temperature in a disk drops with radial distance. 
The region over which this variation occurs is precisely where we expect planetary cores and planets to form in the core accretion scenario and begin their initial stages of migration. 
So far, only a constant adiabatic index of $\gamma=1.4$ has been explored in our previous simulations with radiative transport. 

All of these effects on low-mass planet migration are relevant to understanding whether planet traps can exist, i.e., regions in the disk where a protoplanet could experience zero torque. 
Protoplanets migrating from a smaller radius outwards or from an outer radius inwards would collect at the zero-torque radius, creating areas conducive to planetary mergers, possibly leading to the growth of large cores. 
A planet trap could also be formed by surface density changes, which could create an enhanced feeding zone for these cores \citep{2008A&A...478..929M}; however, it is yet unclear how realistic surface density changes are in disks. 
In contrast, radiation transport might allow traps to exist in even smooth disk structure. 
As outward migration is dependent on the viscosity, disk mass, and adiabatic index, these parameters will influence the radius and breadth of the zero-torque region.
In this work, we present a study on how these effects modify the migration properties of embedded planets. 

The paper is organized as follows: In Section \ref{sec:numerics} we give an overview of our numerical methods. 
We then describe the influence of the adiabatic index, viscosity, and the differences between fixed ortho-to-para and equilibrium gas mixtures on the disk structure in Section \ref{sec:changes}. 
These structural changes influence the migration rate of an embedded $20 M_{\rm Earth}$ planet, which is discussed in Section \ref{sec:influence}. 
In Section \ref{sec:zerotorque} the influence of viscosity and of the different gas mixtures on the zero-torque radius are investigated. 
We then summarize and conclude in Section \ref{sec:summary}.

\section{Numerics and setup}
\label{sec:numerics}

\subsection{Setup}

The protoplanetary disk is modeled as a three-dimensional (3D), non-self-gravitating gas. 
Fluid motion is described by the Navier-Stokes equations, where the equations are solved numerically using a spatially second-order finite volume method that is based on the code {\tt NIRVANA} \citep{1997ZiegYork}. 
The disk is heated solely by internal viscous dissipation, and is allowed to cool by flux-limited diffusion \citep[FLD,][]{1981ApJ...248..321L}.
The FLD approximation allows internally produced energy to diffuse radiatively through the optically thick regions of the disk and into the optically thin regions, where the energy can be radiated away by free-streaming. 
The flux-limiter interpolates between the optically thick and thin regimes and ensures that energy loss never exceeds the free-streaming limit.
 In the code, radiative transport is handled implicitly and it uses the {\tt FARGO} \citep{2000A&AS..141..165M} extension as described in \citet{2009A&A...506..971K}. 
 A more detailed description of the modeling and the numerical methodology is provided in our previous papers \citep{2009A&A...506..971K, 2010A&A.523...A30, 2011A&A...530A..41B}.

The three-dimensional ($r, \theta, \phi$) computational domain (with $266 \times 32 \times 768$ active cells) consists of a complete annulus of the protoplanetary disk centered on the star, extending from $r_{\rm min}=0.4$ to $r_{\rm max}=2.5$ in units of $r_0=a_{\rm Jup}=5.2 AU$. 
For simulations with planets at larger distances from the central star, the computational grid is set to have an inner computational boundary at $0.4 r_P$ and an outer boundary at $2.5 r_P$. 
In the vertical direction the annulus extends $7^\circ$ above the disk's midplane, meaning $83^\circ < \theta < 90^\circ$. 
As we assume that the disk is symmetric for the upper and lower parts, we apply symmetric boundary conditions at $\theta=90^\circ$.
 Here $\theta$ denotes the polar angle of the used spherical polar coordinate system measured from the polar axis. 
 The central star has one solar mass $M_\ast = M_\odot$, and the total disk mass inside [$r_{\rm min}, r_{\rm max}$] is $M_{\rm disk} = 0.01 M_\odot$. 
 The aspect ratio of the disk is calculated self-consistently from the equilibrium structure, given by viscous internal heating and radiative diffusion. This also determines the surface density gradient in the equilibrium state of the disk. 
 To construct the equilibrium state, we first calculate axisymmetric 2D models in the $r$-$\theta$ directions.

The planet is then embedded at $r_P=1.0 a_{\rm Jup}$. 
For the planet's gravity, we use the cubic potential \citep{2006A&A...445..747K, 2009A&A...506..971K} with $r_{\rm sm}=0.5$.
 The planetary potential and its influence on migration is discussed in great detail in \citet{2009A&A...506..971K}. 
 As in our previous simulations we use the opacity law provided by \citet{1985prpl.conf..981L}.

\subsection{2D axisymmetric models}

The initialization through an axisymmetric 2D phase (in the $r-\theta$ plane) reduces the required computational effort substantially. 
The evolution from the initial isothermal state towards the disk equilibrium between viscous heating and radiative transport/cooling (henceforth the disk equilibrium state) takes about 100 orbits, if the disk is started in equilibrium for an isothermal gas and constant H/r. 
The surface density or temperature profiles of the initial (isothermal) state are unimportant as the equilibrium state of the disk solely depends on the disk mass, viscosity, and adiabatic index of the disk.

After reaching the disk equilibrium state, we extend this model to a full 3D simulation by expanding the grid into the $\phi$-direction and by embedding the planet. 
From this starting configuration it is also possible to investigate the vertical structure of an unperturbed disk in the equilibrium state.
This corresponds to a disk with zero total mass flow because the radial boundaries do not allow in- or outflow.

\subsection{Adiabatic index}

\begin{figure}
 \centering
 \resizebox{\hsize}{!}{\includegraphics[width=0.9\linwx]{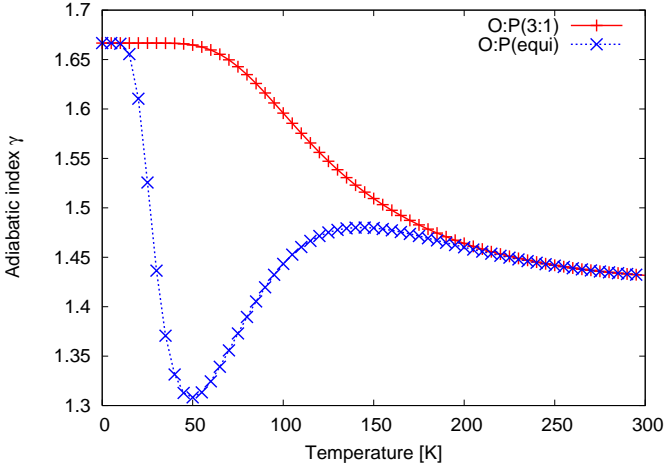}}
 \caption{Adiabatic index for the 3:1 ortho to para mix [O:P(3:1)] and equilibrium gas [O:P(equi)] configuration in dependence of the temperature.
   \label{fig:Adiindex}
   }
\end{figure}

The gas in an accretion disk is primarily molecular hydrogen, which exists as para and ortho-hydrogen for proton spins that are antiparallel and parallel, respectively. 
If enough protonated ions (e.g., $H_3+$) are available to exchange proton spins on timescales that are shorter than the dynamical time, the para- and ortho-hydrogen should be treated as being in statistical equilibrium \citep{2007ApJ...656L..89B}. 
If the dynamical timescale is shorter than the equilibrium timescale, a fixed ratio should be used, where a $3:1$ ortho-to-para mix is common for many astrophysical systems. 
A different mixture of gas changes the ratio of specific heats, which is the adiabatic index $\gamma$. 
The adiabatic index is dependent of the temperature of the underlying gas \citep{2007ApJ...656L..89B}, and astrophysical disks should transition between $\gamma = 5/3$ to $\gamma = 7/5$ {\it near the frost line}. 
This is represented in Fig.~\ref{fig:Adiindex}, where the adiabatic index's dependence on temperature is plotted for the equilibrium (blue) and the $3:1$ ortho-to-para mix (red) gas. 
The adiabatic index can span quite a large region in $\gamma$ for low temperatures. 
To make things clearer for the reader, we will refer to the equilibrium gas mixture as O:P(equi) and as O:P(3:1) for the 3:1 ortho-to-para mix. This is also indicated in Fig.~\ref{fig:Adiindex}. 
In our typical disk simulations (with $\gamma=1.4$) the temperature at $r=1.0 a_{\rm Jup}$ is about $65$K, which is in a region where the adiabatic index changes rapidly, for the O:P(equi) gas and is close to the temperature where the rotational states first activate for the O:P(3:1) gas. 
Such variation could affect an embedded planet's migration rate, as the planet modifies the temperature structure of its surroundings. 
For the disks and orbital times explored in this study, a fixed ortho-to-para ratio is likely the most physical treatment for H$_2$, with the ortho-para equilibration timescale becoming comparable to the orbital timescale outside of 40 AU. 
Nevertheless, we explore both the equilibrium and the 3:1 fixed ratio cases to show limiting behaviors.

\subsection{Viscosity}

The viscosity in an accretion disk is typically assumed to be either constant or an $\alpha$-type viscosity \citep{1973A&A....24..337S}, for which we use here $\nu = \alpha c_s^2 / \Omega$, where
$c_s$ denotes the local adiabatic sound speed and $\Omega$ the local orbital angular velocity. 
If the accretion disk is in a steady state, the viscosity can be determined by the observed mass accretion rate $\dot{M}$ onto the star: 
\begin{equation}
\label{eq:massvisc}
     \dot{M} \approx 3 \pi \nu \Sigma = 3 \pi \alpha c_s^2 / \Omega \, \Sigma,
\end{equation}
where $\dot{M}$ is typically $10^{-9} - 10^{-7} M_\odot/yr$ \citep{1995ApJ...452..736H, 1998ApJ...492..323G} and $\Sigma$ is the surface density. The viscosity determines the structure of the disk and thus influences the migration of embedded planets. Based on models of accretion disks, the observed mass accretion rates constrain $\alpha$ to be roughly between $10^{-4}$ and $10^{-2}$. As will be discussed later in the manuscript, we define the half-disk thickness as
\begin{equation}
\label{eq:height}
              H = \frac{c_{\rm iso}}{\Omega} = \frac{1}{\Omega} \, \sqrt{\frac{P}{\rho}},
\end{equation}
where $c_{\rm iso}$ is the isothermal sound speed, and $P$ and $\rho$ denote the pressure and density in the disk's midplane.

\subsection{Torque calculations}

In previous work, we have discussed the calculation of the torque acting on the planet in great detail. 
Outward migration seems only to be possible when the planet is on a nearly circular orbit \citep{2010A&A.523...A30} and close to the midplane of the disk \citep{2011A&A...530A..41B}. 
Since we want to investigate the occurrence of outward migration and its implications for planet formation and evolution, we assume that the planets remain in the midplane of the disk and are initialized on circular orbits. 
Only the upper half of the disk ($83^\circ \leq \theta \leq 90^\circ$) is needed for the calculation, as the lower half is directly symmetric to the upper half for these conditions.

The torques acting on a $20 M_{\rm Earth}$ planet are calculated to determine the direction of migration. 
A positive torque indicates outward migration, while a negative torque indicates inward migration. 
As the planet is simulated as a point mass, the planetary potential needs to be smoothed. 
We use the cubic potential \citep{2006A&A...445..747K,2009A&A...506..971K} for our calculations:
\begin{equation}
\label{eq:cubic}
\Phi_p^{cub} =  \left\{
    \begin{array}{cc} 
   - \frac{m_p G}{d} \,  \left[ \left(\frac{d}{r_\mathrm{sm}}\right)^4
     - 2 \left(\frac{d}{r_\mathrm{sm}}\right)^3 
     + 2 \frac{d}{r_\mathrm{sm}}  \right]
     \quad &  \mbox{for} \quad  d \leq r_\mathrm{sm}  \\
   -  \frac{m_p G}{d}  \quad & \mbox{for} \quad  d > r_\mathrm{sm} 
    \end{array}
    \right.
\end{equation}
Here $m_P$ is the planetary mass, $d=| \mathbf{r} - \mathbf{r_P}|$ denotes the distance of the disk element to the planet, and $r_{\rm sm}$ is the smoothing length of the potential, usually a fraction of the Hill radius, $R_{\rm H} = (1/3 m_p/M_*)^{1/3}$. 
The construction of the planetary potential is in such a way that, for distances larger than $r_{\rm sm}$, the potential matches the correct $1/r$ potential. For $d < r_{\rm sm}$, the potential is smoothed by a cubic polynomial. 
The parameter $r_{\rm sm}$ is equal to $0.5 R_{\rm H}$ in all our simulations.

The gravitational torques acting on the planet are calculated by summing over all grid cells, which are treated in this case as point masses. We also apply a tapering function to exclude the inner parts of the Hill sphere of the planet \citep{2008A&A...483..325C}. 
This torque-cutoff is necessary to avoid large, probably noisy contributions from the inner parts of the Roche lobe and to disregard material that is possibly gravitationally bound to the planet \citep{2009A&A...502..679C}. 
Here we assume (as in our previous papers) a transition radius of $0.8 R_{\rm H}$.

\section{The equilibrium disk structure}
\label{sec:changes}

As the adiabatic index is connected to the sound speed, a change in the adiabatic index can alter the structure of the disk by modifying the disk viscosity, affecting the disk equilibrium state. 
The effects of these parameters on the disk structure (without an embedded planet) are discussed in this section.

\subsection{Influence of a constant adiabatic index}

\begin{figure}
 \centering
 \resizebox{\hsize}{!}{\includegraphics[width=0.9\linwx]{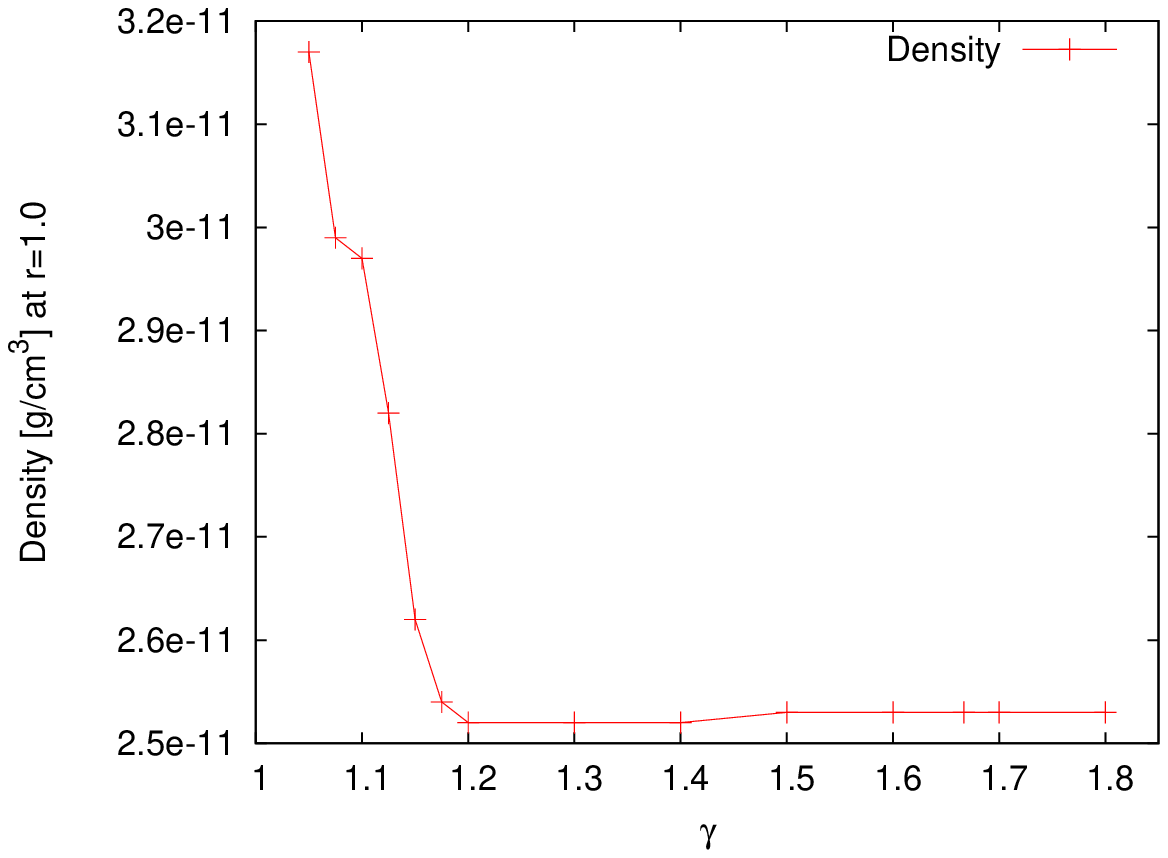}}
 \resizebox{\hsize}{!}{\includegraphics[width=0.9\linwx]{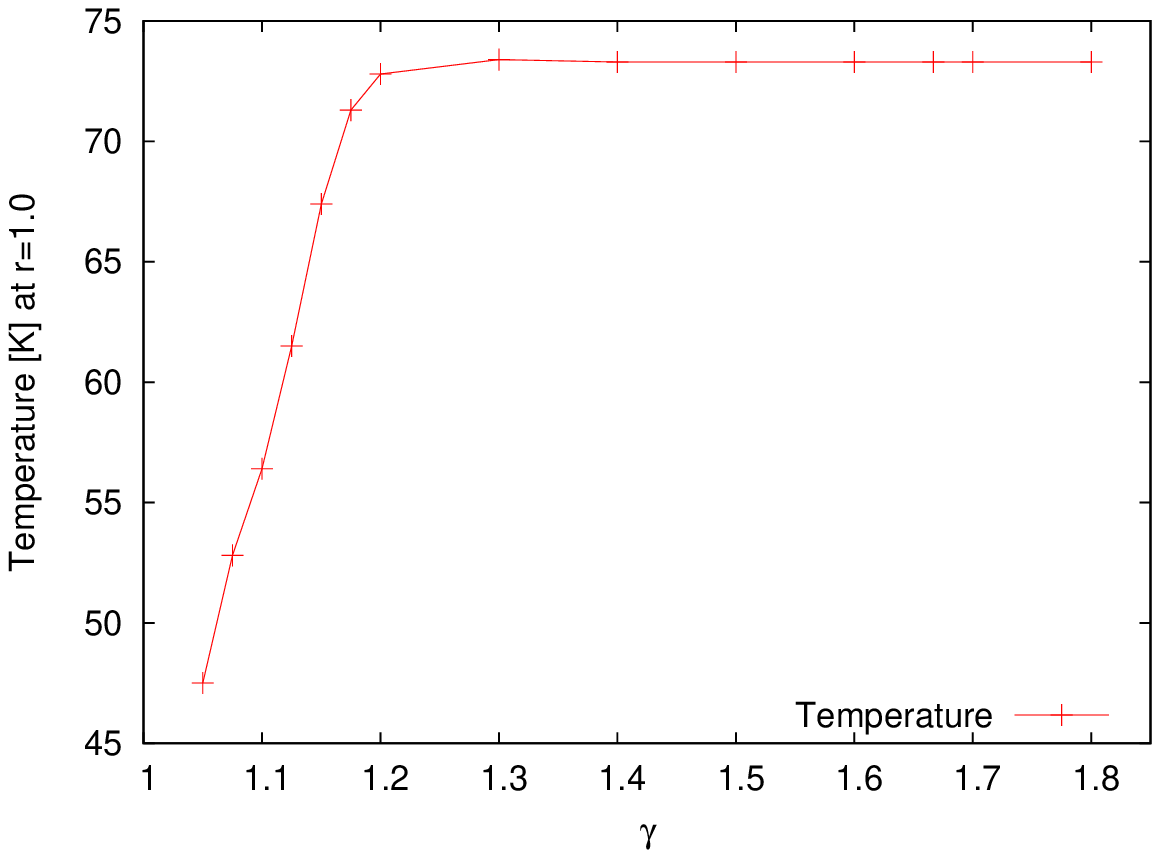}}
 \resizebox{\hsize}{!}{\includegraphics[width=0.9\linwx]{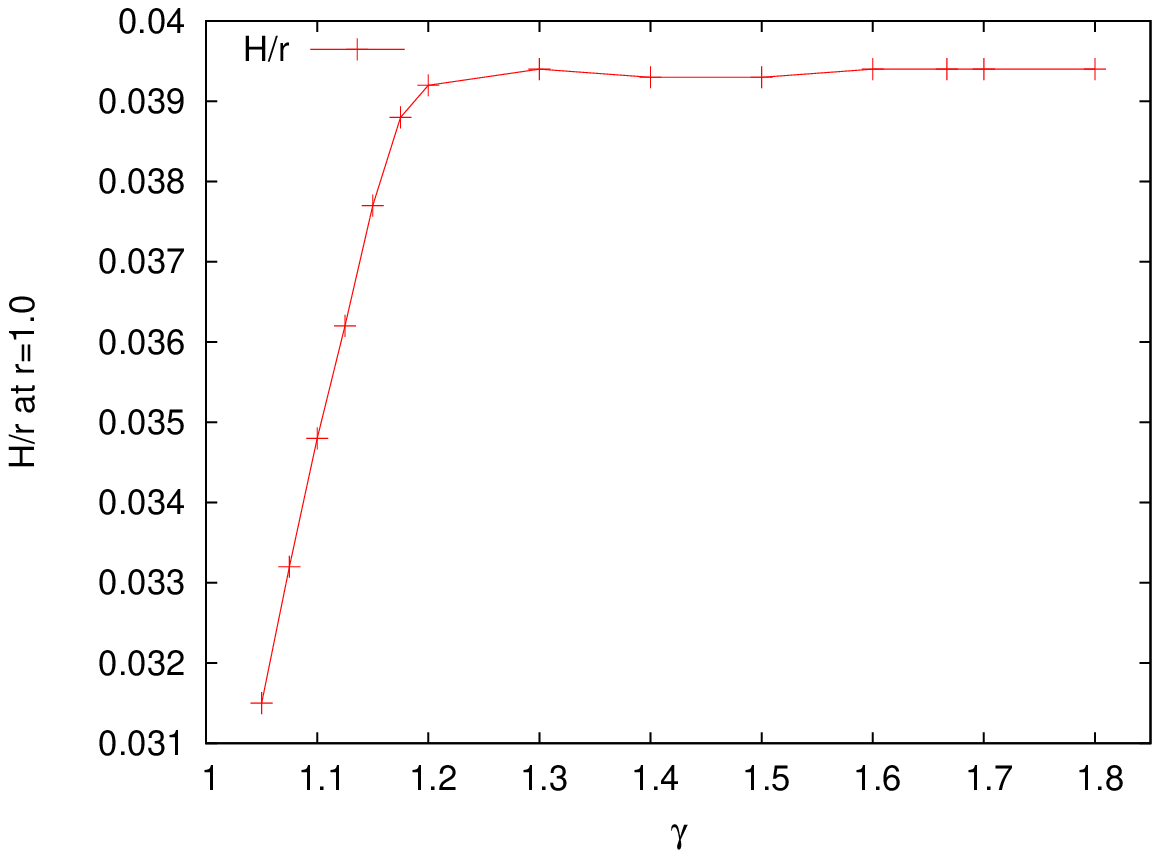}}
 \caption{Densities (top), temperatures (middle), and aspect ratios (bottom) from the 2D axisymmetric disks with different (constant) adiabatic indices, taken after the disks have reached  their equilibrium state. 
Each quantity is evaluated using midplane values of the disk at a radius $r=1 a_{\rm Jup}$. The kinematic viscosity has been set to $\nu = 10^{15}$cm$^2$/s for all models.
   \label{fig:AdiRTHR}
   }
\end{figure}

For the following simulations, we limit ourselves to constant values of $\gamma$ in the range of $1.05 \leq \gamma \leq 1.8$, which is larger than the range of the adiabatic index in Fig.~\ref{fig:Adiindex}. 
We include $\gamma$ values smaller than $1.3$ to capture the transition to isothermal disks, and higher $\gamma$ values to capture additional possible stiffness resulting from magnetic fields.
In a first set of simulations we study disks with a constant kinematic viscosity of $\nu = 10^{15}$\,cm$^2$/s. 
For this case, we do not expect a change in the disk structure for different values of $\gamma$ because the disk equilibrium  state is determined by a balance between viscous heating and radiative cooling, both of which do not depend of $\gamma$. 
Indeed, as can be seen in Fig.~\ref{fig:AdiRTHR}, for a wide range of $\gamma$, the midplane density and temperature and the resulting aspect ratio at $r=1.0 a_{\rm Jup}$ for a 2D axisymmetric disk do not depend on $\gamma$. 
Only for small values of $\gamma$ is the disk structure different.
The reason for this change is the onset of convection for $\gamma \lsim 1.2$.
As the adiabatic index is lowered, the gas becomes more compressible owing to the larger degrees of freedom that must be excited before the thermodynamic temperature can be raised. This leads to a higher midplane density, a lower midplane temperature, and a lower aspect ratio (see eq.~\ref{eq:height}) in the disk at $r=1.0 a_{\rm Jup}$.

%
Radial density and temperature gradients result in vortensity and entropy ($S \propto \frac{p}{\rho^\gamma}$) gradients. 
These  vortensity and entropy gradients across the horseshoe region determine the magnitude and sign of the corotation torque, which, if positive, can lead to a total positive torque, indicating outward migration in fully radiative disks. 
As the density decreases and the temperature increases with increasing $\gamma$ (for $\gamma<1.2$), the strength of the gradient in entropy might change and therefore influence planetary migration. 
The influence of this effect might be stronger for the O:P(equi) gas disk, as the adiabatic index can become lower compared to the O:P(3:1) gas disk.


\begin{figure}
 \centering
 \resizebox{\hsize}{!}{\includegraphics[width=0.9\linwx]{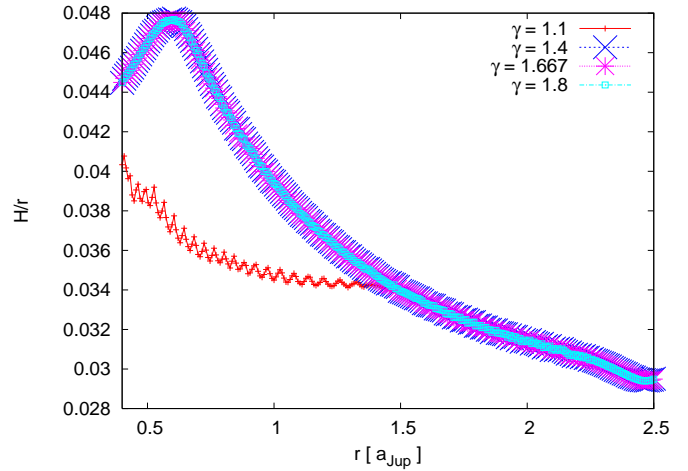}}
 \caption{Aspect ratio ($H/r$) of a fully radiative disk with different adiabatic indices without an embedded planet. The aspect ratio was computed using midplane values of the constant viscosity disks (see eq.~\ref{eq:height}).
   \label{fig:Hrgammaall}
   }
\end{figure}

The radial variation of the aspect ratio of a fully radiative disk with different adiabatic indices is shown in Fig.~\ref{fig:Hrgammaall}.

For $\gamma=1.1$, the aspect ratio is smaller at $r<1.5 a_{\rm Jup}$ and the fluctuations in this region are a sign of convection. 
However, such a small adiabatic index is not predicted for accretion disks, as seen by the $\gamma (T)$ profile in Fig.~\ref{fig:Adiindex}, unless efficient heating and cooling push the disk toward isothermal behavior or the gas is undergoing  dissociation or ionization.
In contrast, the profiles for a higher adiabatic index show an overall drop in aspect ratio with increasing distance to the central star (after a small increase for smaller distances to the central star, $r<0.7$). 
Such a feature in the disk's structure is the opposite of what one would expect. 
When taking stellar irradiation into account an increasing $H/r$ profile for larger radii may be possible, e.g. \citet{2004A&A...417..159D}. However, the disk's opacity can greatly influence the $H/R$ profile.  

With increasing distance from the central star, the temperature is decreasing. 
For temperatures less than $T\approx 155$ K, the opacity increases with increasing temperature. 
At higher temperature, the opacity decreases. 
So, with increasing distance to the central star, the opacity first increases and, when $T \approx 155 K$, the opacity decreases. 
This is exactly the point in the disk ($r\approx 0.7$) where the aspect ratio $H/r$ begins to decrease. 
Please note that the height $H$ of the disk is still increasing, but at a lower rate, when $T < 155$ K compared to $T > 155$ K. 
Hence, the structure of the disk is a result of the opacity law that is used in these calculations.

The $H/r$ profiles for the disks with $\gamma>1.4$ are identical, as the disk equilibrium state is determined by a balance of viscous heating and radiative cooling, neither of which depends on $\gamma$, at least for the constant $\nu$ assumed for these disks.


\subsection{Influence of viscosity}
\label{subsec:disk-visc} 
\begin{figure}
 \centering
 \resizebox{\hsize}{!}{\includegraphics[width=0.9\linwx]{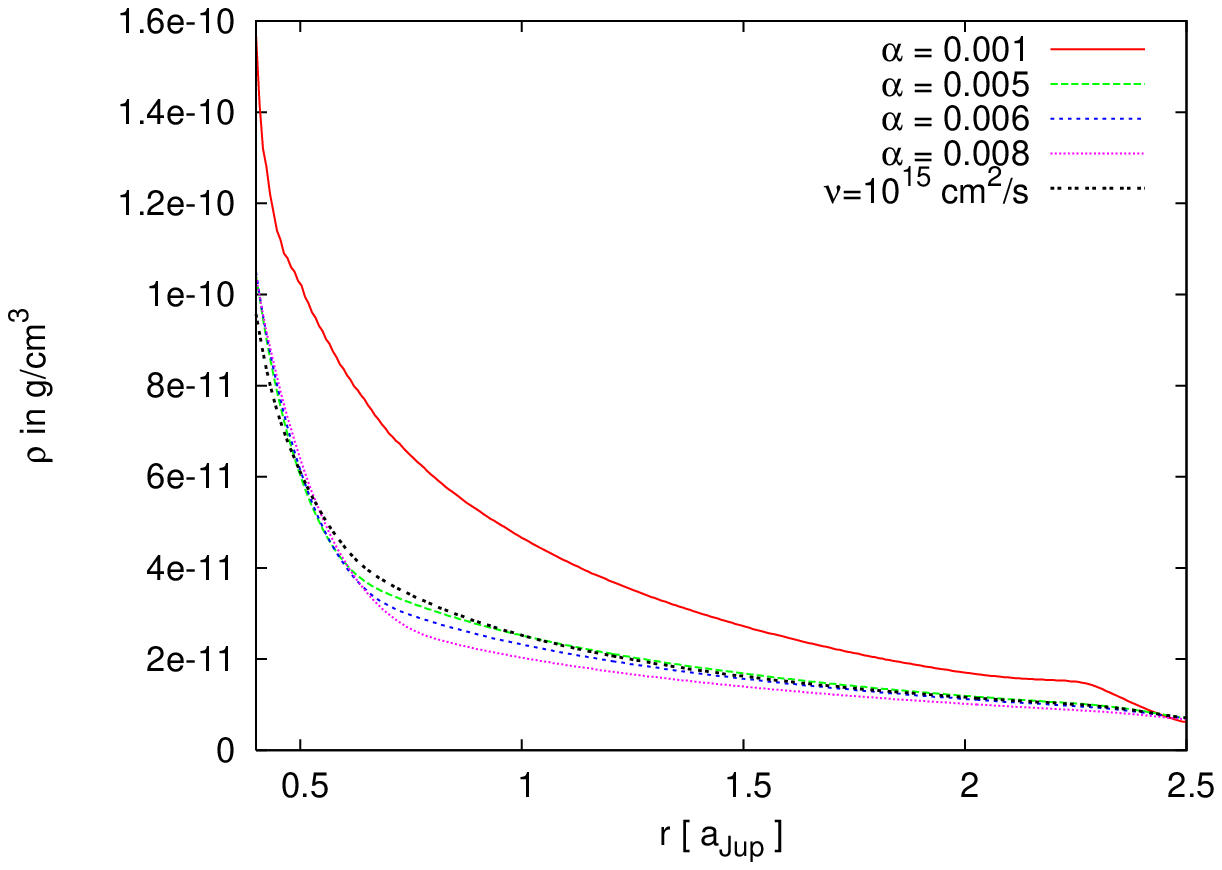}}
 \resizebox{\hsize}{!}{\includegraphics[width=0.9\linwx]{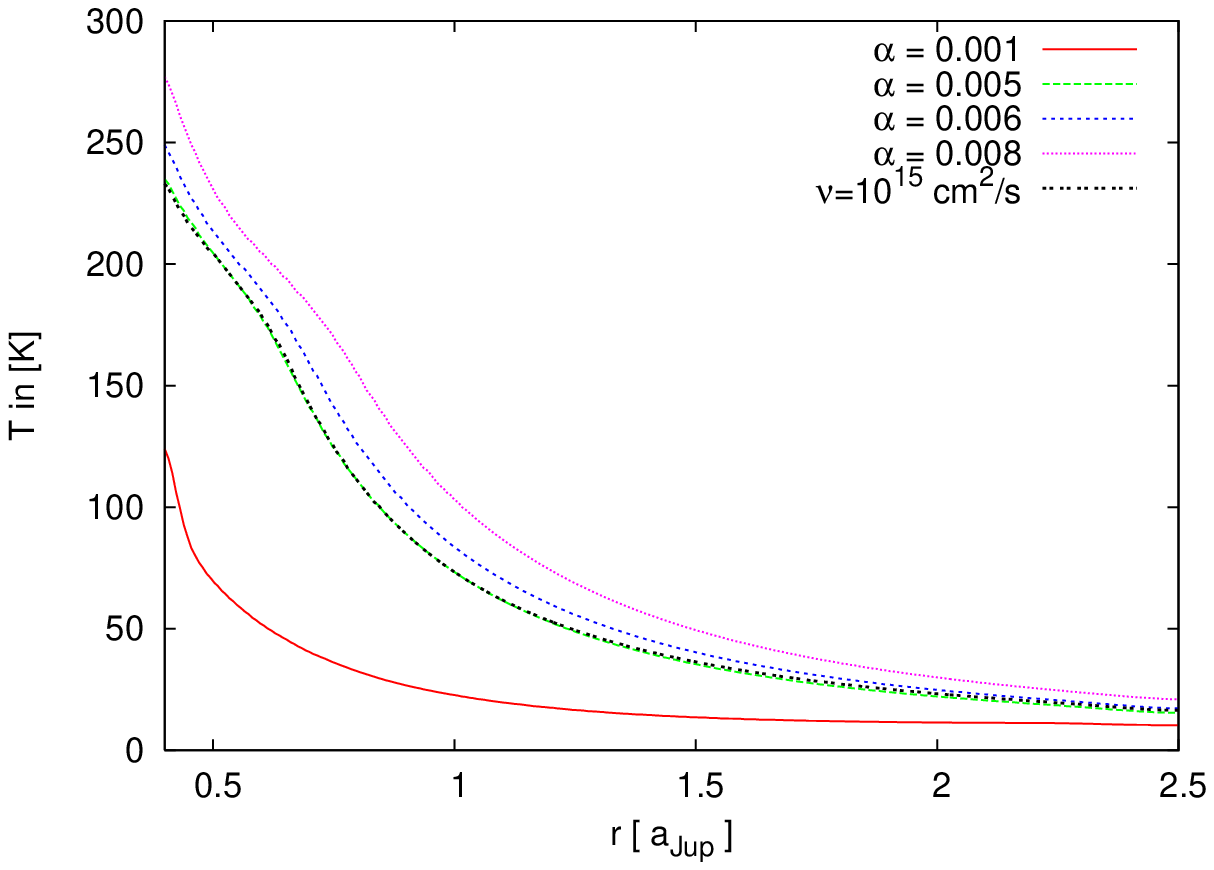}}
 \caption{Density (top) and temperature (bottom) profiles of the fully radiative disks with different viscosities, all taken in the disk equilibrium state. Densities and temperatures are measured in the midplanes of the disks.
   \label{fig:RhoTemp2D}
   }
\end{figure}

As can be seen in Fig.~\ref{fig:RhoTemp2D}, where we display the midplane density (top) and temperature (bottom) of fully radiative disks in the disk equilibrium state, a change in the viscosity affects the disk's equilibrium structure, altering the density and temperature profiles of the disk.
In this figure we show the densities and temperatures for $\alpha$-viscosities  $0.001 \leq \alpha \leq 0.008$ and for a constant viscosity of $\nu = 10^{15}$\,cm$^2$/s, a value used in our previous simulations \citep{2009A&A...506..971K, 2010A&A.523...A30, 2011A&A...530A..41B}. 

For increasing $\alpha$-viscosity, the density in the midplane of the disk becomes smaller, but the differences in density become smaller for higher viscosities. 

For the temperature profile, the trend is reversed: a higher viscosity results in a higher temperature in the midplane of the disk. As the viscosity increases, so does the viscous heating. 
The heating process mainly takes place in the midplane of the disk, as the viscosity is highest there. 
Not only is the temperature higher in the midplane, but the temperature profiles also change with viscosity. 
For our constant viscosity of $\nu = 10^{15}$\,cm$^2$/s, the radial slope of temperature was $\propto r^{-1.8}$ at $r =1 a_{\rm Jup}$, while it now depends on the chosen $\alpha$-value. The slope of temperature $\beta_{\rm T}$, where
$T \propto r^{-\beta_{\rm T}}$, is displayed in Table~\ref{tab:temptorq2D} for $r=1 a_{\rm Jup}$.

\begin{table}[b]
 \centering
  \begin{tabular}{|l|l|l|}\hline
  $\alpha$ & $\beta_{\rm T}$ & $H (r=1.0 a_{\rm Jup})$\\\hline
  $0.001$ & $1.60$ & 0.0219 \\\hline
  $0.002$ & $1.70$ & 0.0277 \\\hline
  $0.003$ & $1.75$ & 0.0324 \\\hline
  $0.004$ & $1.80$ & 0.0363 \\\hline
  $0.005$ & $1.80$ & 0.0393 \\\hline
  $0.006$ & $1.85$ & 0.0419 \\\hline
  $0.007$ & $1.87$ & 0.0442 \\\hline
  $0.008$ & $1.90$ & 0.0468 \\\hline
  $\nu=const.$ & $1.80$ & 0.0395 \\\hline
  \end{tabular}
 \caption{The slope of the temperature profile $\beta_{\rm T}$, where $T \propto r^{-\beta_{\rm T}}$, and the disk thickness $H$ as a function of disk viscosity.
    The quantities are evaluated at disk radius $r=1.0 a_{\rm Jup}$.
 \label{tab:temptorq2D}
 }
\end{table}

The temperature slope has a direct influence on the migration of embedded planets, e.g., \citet{2010MNRAS.401.1950P, 2011MNRAS.410..293P}, as the Lindblad torque and the entropy related corotation torque are very sensitive to it. 
A change in $\beta_{\rm T}$ therefore influences the torque acting on an embedded planet quite severely \citep{2011arXiv1103.3502A}. 

The aspect ratio $H/r$ of the disk also increases for increasing viscosity (see Table~\ref{tab:temptorq2D}), indicating another factor that influences planetary migration in these disks. 
For isothermal disks, this has been know for a while \citep{2002ApJ...565.1257T}.

The density, temperature and $H/r$ profiles of our constant viscosity simulation match quite well with the $\alpha=0.005$ simulations, indicating that one could expect a very similar torque acting on a planet embedded in these disks. 
However, as the $\alpha$-viscosity is not constant in $r$, one might also expect some differences in the spiral wave densities and temperatures in these disks. In addition, the area near the planet might be subject to some changes for different viscosities.

\subsection{Influence of a varying adiabatic index}

\begin{figure}
 \centering
 \includegraphics[width=0.9\linwx]{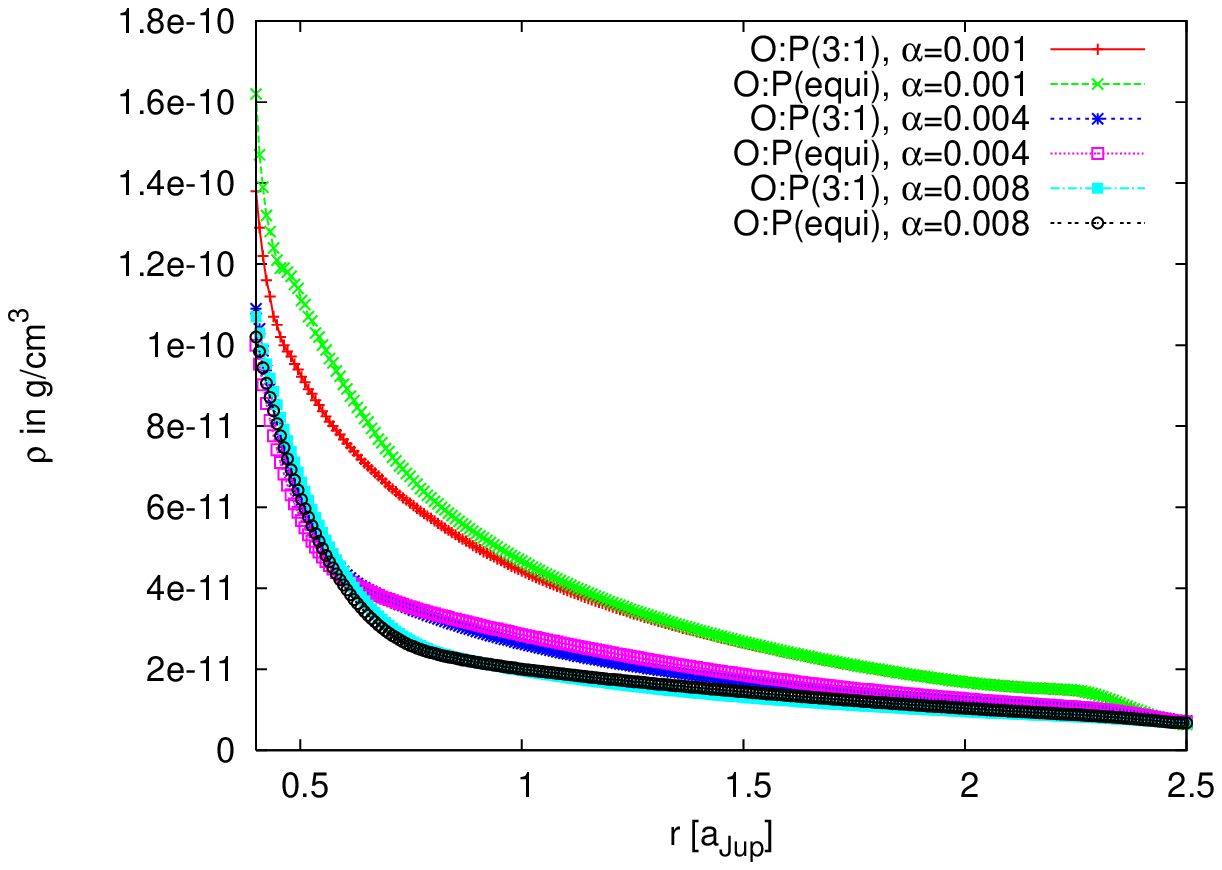}
 \includegraphics[width=0.9\linwx]{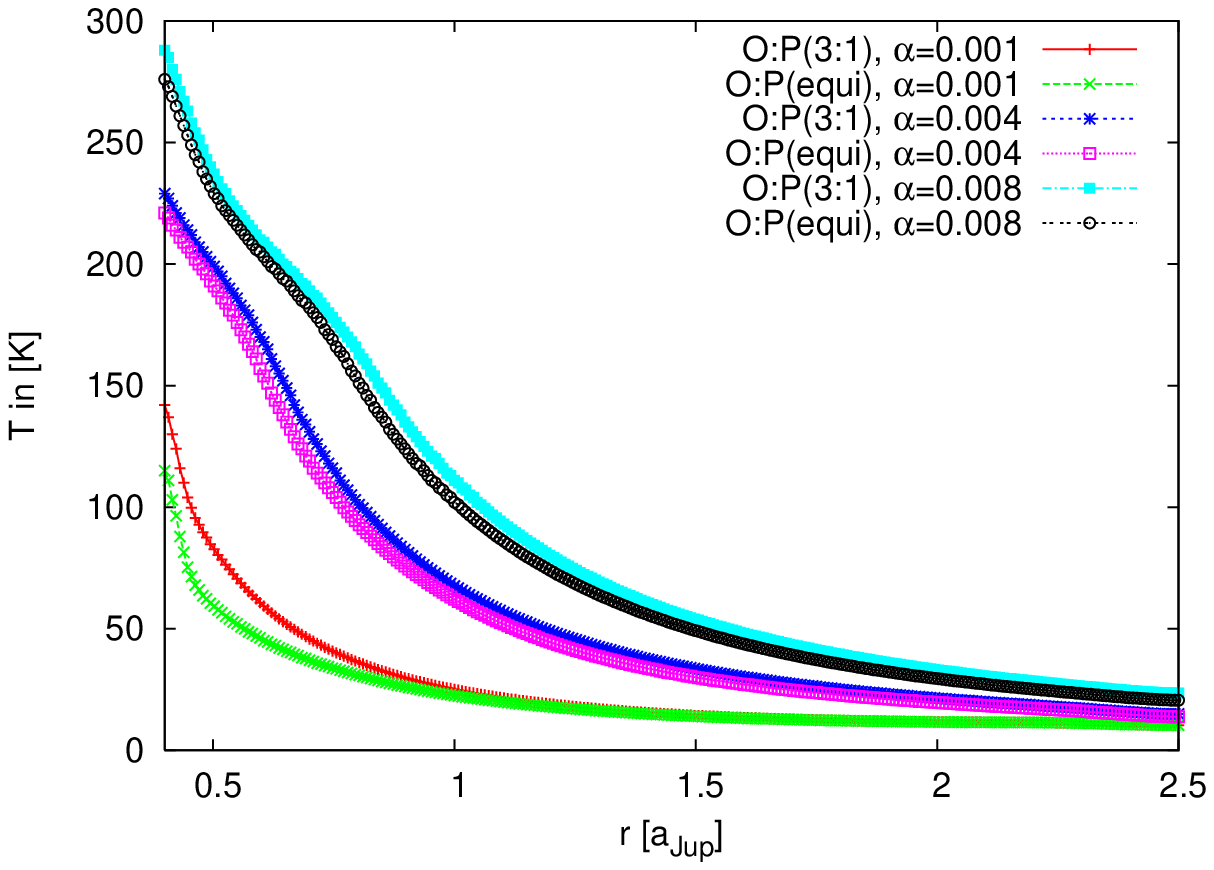}
 \caption{Midplane density (top) and temperature (bottom) profiles for the $\alpha$ viscosity disks with different assumptions for the ortho-para ratio of the gas.
   \label{fig:RTadialpha}
   }
\end{figure}

In Fig.~\ref{fig:RTadialpha} the temperature (bottom) and density (top) profiles in the midplane of fully radiative disks are displayed. 
The disks feature different $\alpha$ viscosities and two different ortho-para mixtures of the gas [O:P(equi) and O:P(3:1)]. 
As expected from simulations with a constant adiabatic index, an increasing $\alpha$ viscosity results in lower densities and higher temperatures in the midplane of the disk.

The gradients in density and temperature are comparable to the gradients for constant adiabatic indices. 
It also seems that the mixture of the gas does not have a huge influence on the temperature and density profiles for large viscosities. 
However, for $\alpha=0.001$ there are differences in the density and temperature profiles between disks that have different ortho-para ratios. 
For $r < 1.0 a_{\rm Jup}$, the temperature for the disk equilibrium state is lower in the O:P(equi) than in the O:P(3:1). 
As a result, the density is higher in the O:P(equi) disk. 
As $\gamma$ is largely reduced for temperatures around $T \approx 50$K for ortho-para ratios O:P(equi), this behavior is consistent with the results of the constant $\gamma$ simulations.

\begin{figure}
 \centering
 \resizebox{\hsize}{!}{\includegraphics[width=0.9\linwx]{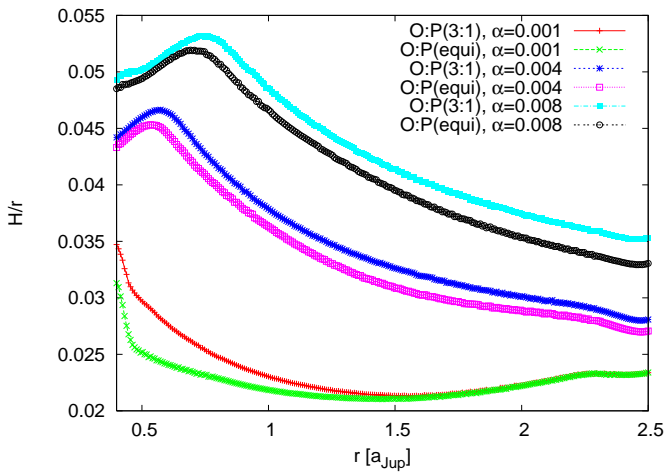}}
 \caption{Aspect ratio $H/r$ of the accretion disks for different $\alpha$ viscosities and different assumptions for the ortho-para ratio of the gas.
 \label{fig:HRadialpha}
   }
\end{figure}

In Fig.~\ref{fig:HRadialpha} the aspect ratios $H/r$ for the same disks are shown. The profiles are very similar to those in disks with a constant adiabatic index and an $\alpha$ viscosity. 
In the $\alpha = 0.001$ case, the aspect ratio profile is smaller and approximately constant.
For the more viscous disks, the aspect ratio increases for small $r$, but then decreases with increasing $r$ continuously. A higher viscosity leads to a higher aspect ratio in the disk. The O:P(equi) shows a slightly smaller aspect ratio for viscosities $\alpha \geq 0.004$ compared to the O:P(3:1).
The aspect ratio differs by about $10 \%$ for the high viscosity cases at $r=1.0 a_{\rm Jup}$. 
This difference at this location can influence the torque acting on embedded planets, as seen in isothermal disks.

A higher disk viscosity leads to a larger temperature in the midplane. 
Recently, \citet{2011A&A...536A..77B} have shown that in fully radiative disks with viscous heating, a convective region arises near the central star, where its radial extent depends on disk mass. 
A higher disk mass resulted in a larger convective zone in the disk. 
The density and temperature structures of the disk with constant viscosity seem to correspond to a disk with $\alpha \approx 0.005$ (see Fig.~\ref{fig:RhoTemp2D}).

\begin{figure}
 \centering
 \resizebox{\hsize}{!}{\includegraphics[width=0.805\linwx]{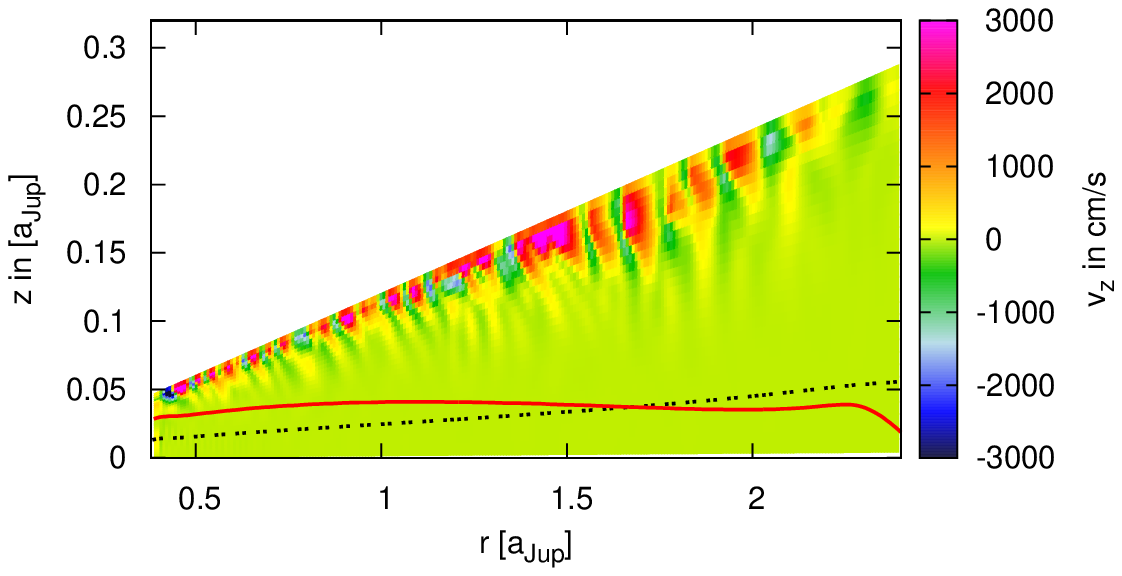}}
 \resizebox{\hsize}{!}{\includegraphics[width=0.805\linwx]{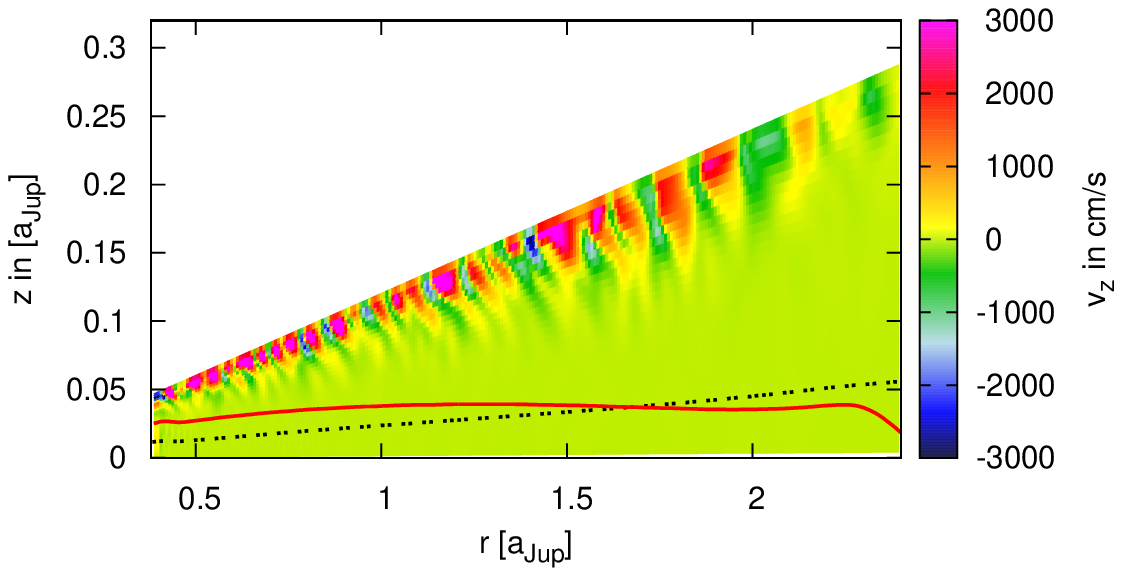}}
 \resizebox{\hsize}{!}{\includegraphics[width=0.805\linwx]{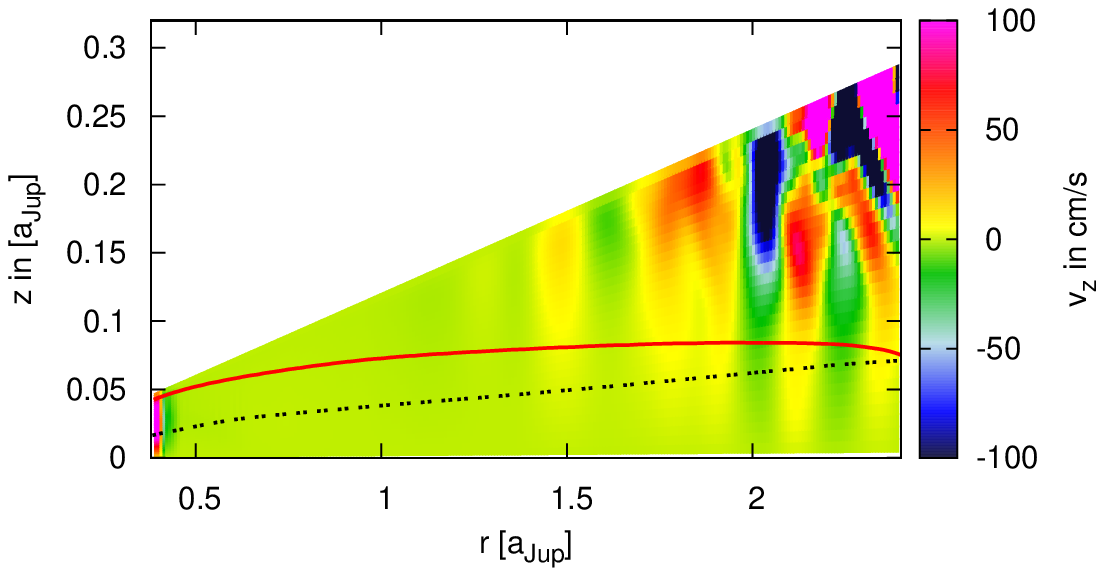}}
 \resizebox{\hsize}{!}{\includegraphics[width=0.805\linwx]{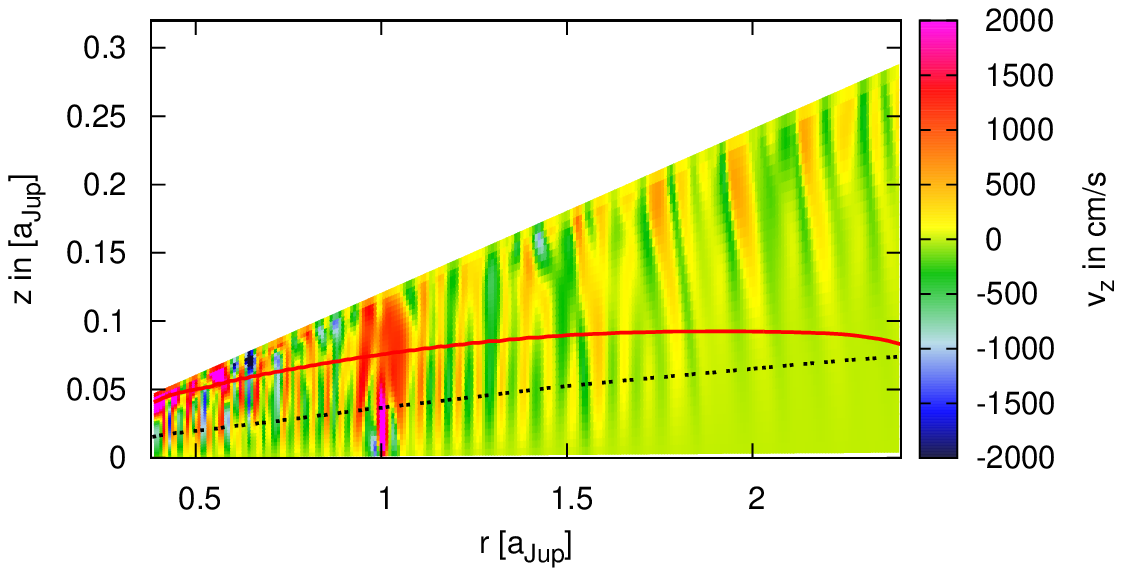}}
 \caption{Velocities in $z$-direction for disks with $\alpha=0.001$  and O:P(3:1) (top), $\alpha=0.001$ and O:P(equi) (2nd from top), $\alpha=0.004$ and O:P(equi) (3rd from top), and a fixed $\gamma=1.1$ with constant viscosity (bottom). 
 The black line indicates $H(r)$ in the disk and the red line delineates  $\tau=1.0$ (the optical depth integrated from the top of the grid toward the midplane of the disk). 
 Please note the different color scales and that they only show a subset of the range of values to better highlight the internal structure.
   \label{fig:2D001vxy}
   }
\end{figure}

We display in Fig.~\ref{fig:2D001vxy} the velocity in the $z$-direction for the $\alpha=0.001$ disks with two different ortho-para gas mixtures, for the $\alpha=0.004$ and O:P(equi) disk, and for the fixed $\gamma=1.1$ disk. 
In black we indicate $H(r)$ and in red the $\tau=1.0$ line in the disk (integrated from the top of the grid toward the midplane). 
We note that when taking both sides of the disk into account, convective eddies cross the midplane. 
However, \citet{2011A&A...536A..77B} have shown that the first signs of convection can also be seen in simulations where only one half of the disk is simulated. 
The extent to which the convective region inside the disk depends on the adiabatic index should be tested in simulations of disk instabilities. 
As we want to focus here on the influence of the adiabatic index on planet migration, we do not pursue this topic further. 
We do note that \citet{1991ApJ...375..740R} have shown that a lower adiabatic index can increase the susceptibility of the disk to convection.

For the $\alpha=0.001$ disk with either the O:P(3:1) or the O:P(equi) gas, variations in the vertical velocity of the gas in the disk's upper layers can be seen in Fig.~\ref{fig:2D001vxy}.  
However, these fluctuations are limited to the low-density material, and do not reach all the way down to the high optical depth region. 
Convection does not appear to be present. 
Fluctuations such as those in the top two panels of Fig.~\ref{fig:2D001vxy} only occur for the thinner disks (small $H/r$) that have a very large vertical range in density, which is difficult to resolve numerically (in this case no-density floor is used, possibly increasing the effect). 
\citet{2012arXiv1209.2753N} stated that vertical instabilities in isothermal discs can arise, if the viscosity is negligible. 
This vertical shear instability can then drive a non negligible transport of angular momentum through Reynolds stress with a the corresponding viscous alpha value $\alpha=0.001$, which is the value of viscosity assumed in the simulations shown in the top two plots in Fig.~\ref{fig:2D001vxy}, indicating that the shown fluctuations might be resulting from the vertical shear instability. 
However, detailed studies of the vertical shear instability for the case of fully radiative disks is beyond the scope of the presented work.

Convection, on the other hand, can only be present in the optically thick regions of the disk, where radiative cooling is inefficient. For larger viscosities ($\alpha=0.004$), the fluctuations in the upper layers mostly vanish, except in the low-density region at large $r$, and are totally diminished for $\alpha=0.008$ (not shown). Large viscosities suppress the growth of vertical oscillations, in agreement with the vertical shear instability \citep{2012arXiv1209.2753N}.

On the other hand, the fixed $\gamma=1.1$ disk shows fluctuations reaching all the way down to midplane, where the disk is optically thick. 
Please also note here that the absolute values of the velocities are larger for the $\gamma = 1.1$ disk compared with the other presented simulations. 
In this case, we have to check if convection is really present in the disk, as we already suspected from the aspect ratio profile of the disk (see Fig.~\ref{fig:Hrgammaall}).

In order to test whether convection should be present in the disk, we compute the adiabatic gradient as follows:
\begin{equation}
\label{eq:adigrad}
\beta = \frac{d \ln T}{d \ln p} \ .
\end{equation}
The adiabatic gradient is evaluated at a two-degree angle from the midplane, which roughly corresponds to the half opening angle of the optical thick parts of the disk.

\begin{figure}
 \centering
 \resizebox{\hsize}{!}{\includegraphics[width=0.805\linwx]{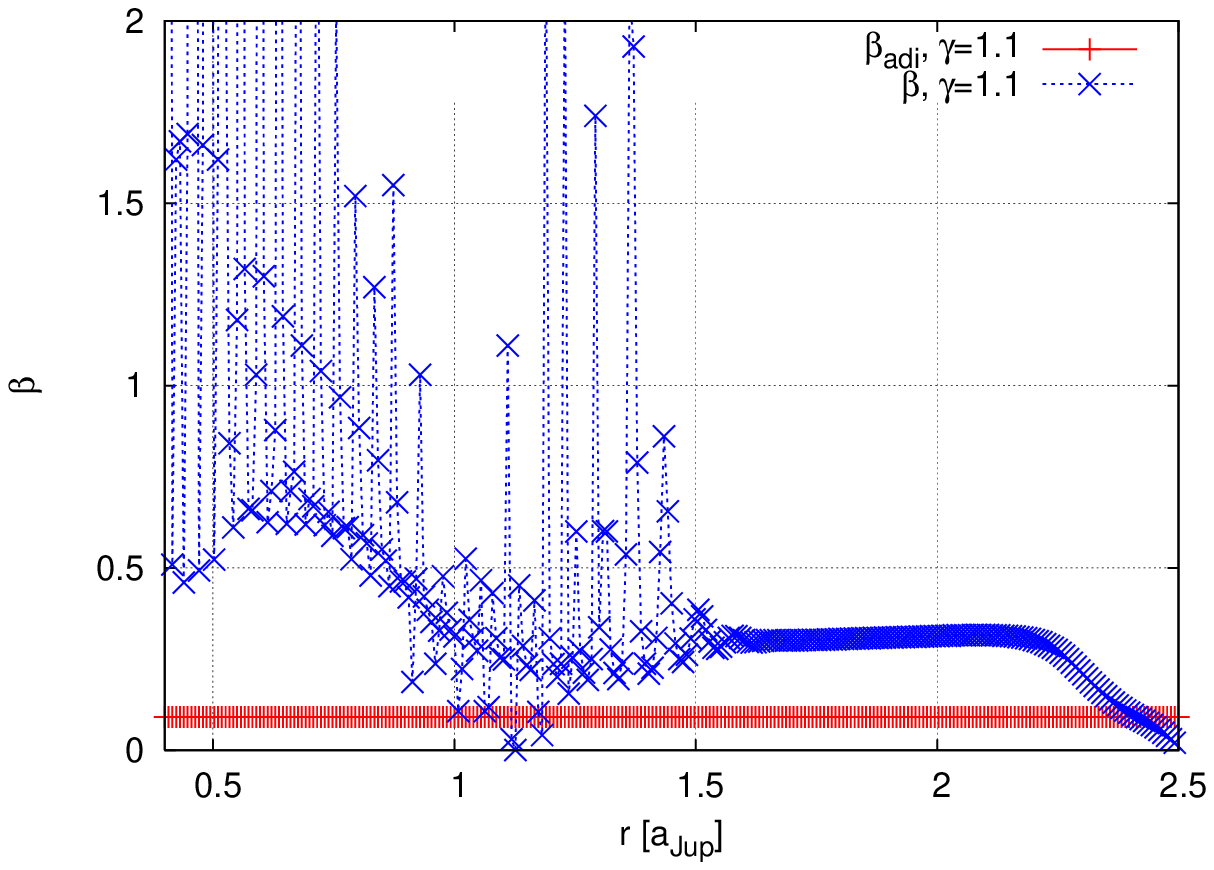}}
 \resizebox{\hsize}{!}{\includegraphics[width=0.805\linwx]{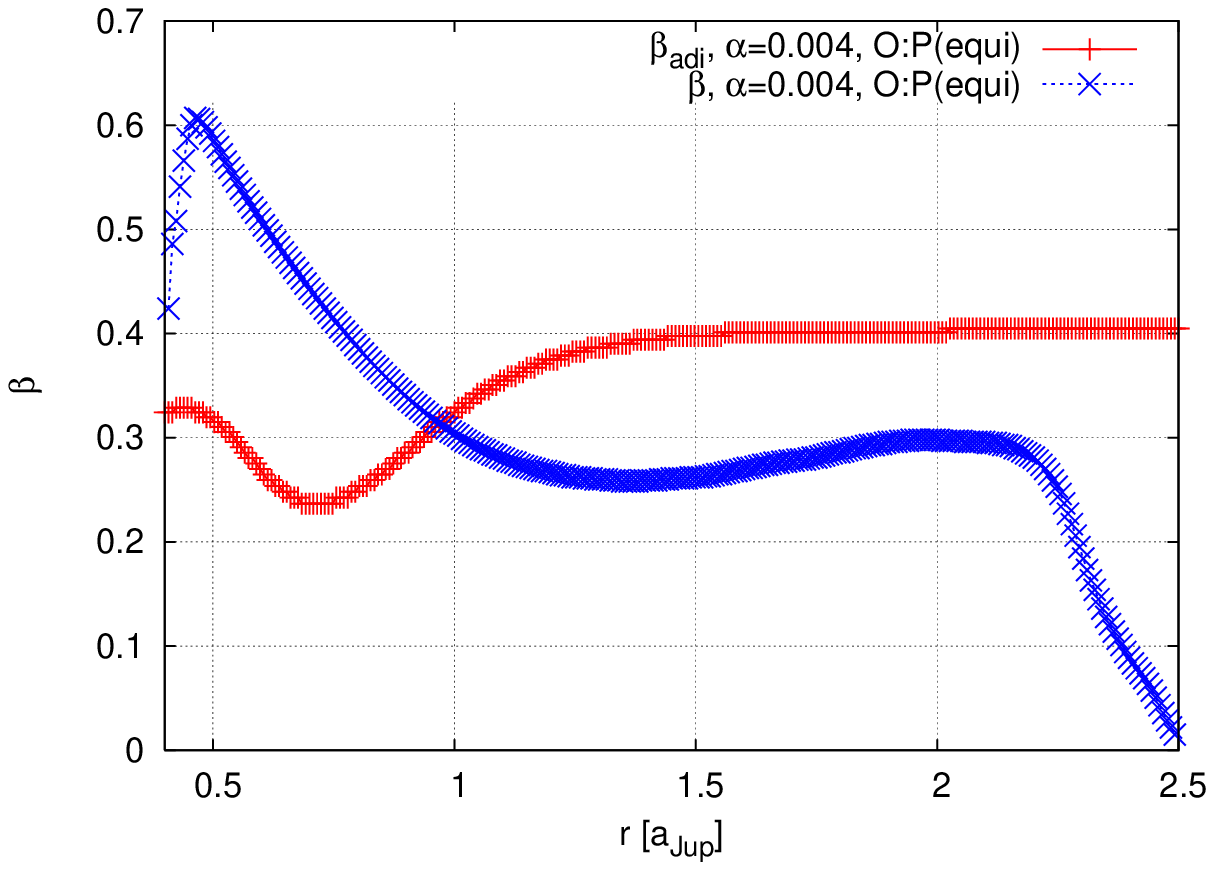}}
 \caption{Adiabatic gradients for the $\gamma=1.1$ disk with constant viscosity (top) and for the $\alpha=0.004$ disk with O:P(equi) (bottom). The gradients have been evaluated in the optical thick regions of the disk at an angle of $2^\circ$ from the midplane.
   \label{fig:Adigrad}
   }
\end{figure}

For $\beta > \beta_{\rm adi}$, convection can be present in the disk, where $\beta_{\rm adi}$ is defined as
\begin{equation}
\label{eq:adiabat}
\beta_{\rm adi} = \frac{\gamma_{\rm local}-1}{\gamma_{\rm local}} \ .
\end{equation}
Here, $\gamma_{\rm local}$ represents the local adiabatic index, which in our case, is the local adiabatic index along the two-degree angle of the disk.
 In Fig.~\ref{fig:Adigrad} we present the the adiabatic gradients of the $\gamma=1.1$ disk (bottom in Fig.~\ref{fig:2D001vxy}) and the $\alpha=0.004$ disk with O:P(equi) (3rd from top in Fig.~\ref{fig:2D001vxy}).

In the case of $\gamma=1.1$, $\beta > \beta_{\rm adi}$, indicating that convection should be present in the disk, which corresponds to the vertical fluctuations for this simulation in the optically thick regions of the disk. 
The strong fluctuations in $\beta$ indicate that the disk is highly unstable. 
Please also note that in the inner regions of the disk, the difference between $\beta$ and $\beta_{\rm adi}$ is about a factor of $5$. 
For the $\alpha=0.004$ disk, only the inner regions feature $\beta > \beta_{\rm adi}$. 
However, fluctuations in the velocity have not been observed (Fig.~\ref{fig:2D001vxy}). 
In addition, the difference between $\beta$ and $\beta_{\rm adi}$ is at maximum a factor of $2$, which is much lower than in the $\gamma=1.1$ case. 
If convection is present in the $\alpha = 0.004$ disk, then it must be much less vigorous than seen in the $\gamma = 1.1$ disk.

\section{Influence on planetary migration}
\label{sec:influence}

Because the above results show that the adiabatic index and the viscosity affect disk profiles, we also expect that they will change the migration rates of embedded planets. 
Embedding a small mass planet in such a disk will, of course, change the profiles of the disk. 
However, for most viscosities and adiabatic indices, the changes that the low-mass planet will cause are limited to the vicinity of the planet, so that the overall structure of the disk remains intact. 
To determine the direction of migration of protoplanets in disks, we embed $20 M_{\rm Earth}$ planets in our disks and measure the torque acting on those planets. 
A positive torque represents outward migration, while a negative torque indicates inward migration. 
The torque acting on planets on circular, non-inclined orbits is directly proportional to the migration rate. 
In cases where the planet is embedded in the convective region in the disk (for $\gamma \lsim 1.1$), the torques have been averaged over $20$ planetary orbits.

\subsection{Constant adiabatic index}
\label{subsec:CAI}

\begin{figure}
 \centering
 \resizebox{\hsize}{!}{\includegraphics[width=0.9\linwx]{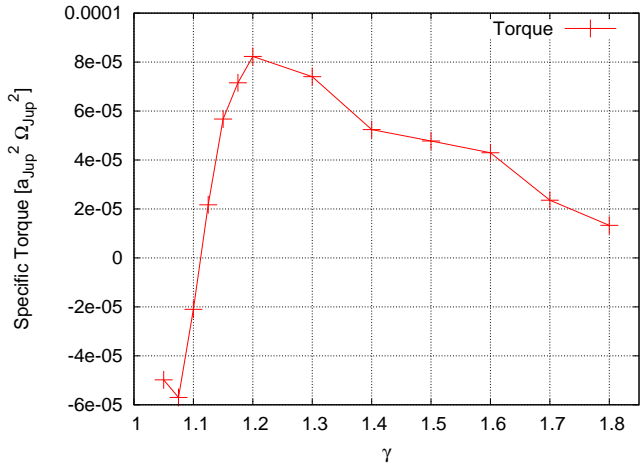}}
 \caption{The total torque $\Gamma_{\rm tot}$ acting on a $20 M_{\rm Earth}$ planet embedded at $r=1$ in $\nu = 10^{15}$\,cm$^2$/s disks for different (constant) adiabatic indices. 
 The torques have been measured in the equilibrium state of the disk.
   \label{fig:AdiTorque}
   }
\end{figure}

In Fig.~\ref{fig:AdiTorque} we show the total torque $\Gamma_{\rm tot}$ acting on planets that are on circular orbits and are embedded in disks that have different adiabatic indices.  The viscosity is held constant at $\nu = 10^{15}$\,cm$^2$/s. 
The torque acting on the planet is positive when the adiabatic index is larger than $1.12$ , while it is negative, indicating inward migration, for $\gamma < 1.12$. 
The maximum of the torque is at $\gamma=1.2$, and the torque decreases for lower and higher $\gamma$'s. 
The most negative torque is at $\gamma=1.075$.

\begin{figure}
 \centering
 \resizebox{\hsize}{!}{\includegraphics[width=0.9\linwx]{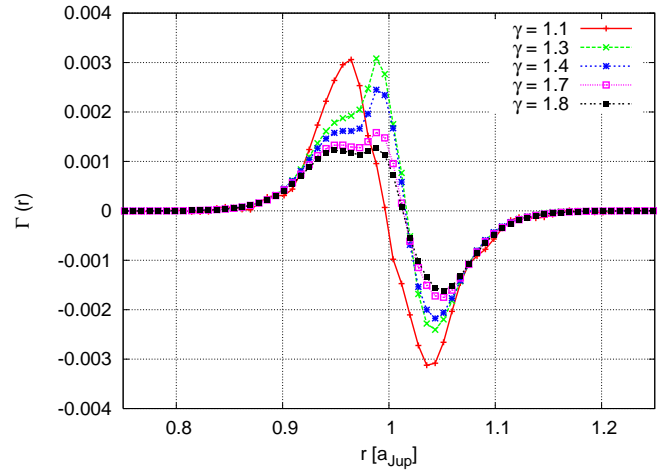}}
 \caption{Profiles of the radial torque density $\Gamma (r)$ acting on $20 M_{\rm Earth}$ planets embedded in disks with different adiabatic indices (during the disk equilibrium state). 
 In contrast to the total torque, the radial torque density has not been averaged in time.
   \label{fig:AdiconstGamma}
   }
\end{figure}

For $\gamma \geq 1.12$, changing the adiabatic index influences the overall results of planetary migration only by magnitude, not by direction. 
At high adiabatic indices, the torque acting on the embedded planet is decreased by a factor of $3$ to $4$ compared to the maximum torque at $\gamma=1.2$. The tendency of outward migration, however, remains intact. 
An increase in $\gamma$ leads to an increase in the adiabatic sound speed, which in turn decreases the amplitude of the total torque. 
It also reduces the amplitude of the (negative) entropy gradient, which then reduces the amplitude of the (positive) corotation torque. 
This can be seen by the decrease of the (positive) spike in the torque density distribution (see Fig.~\ref{fig:AdiconstGamma}). 
For lower adiabatic indices, the direction of migration is reversed as $\gamma$ approaches the value for isothermal disks.
 However, the torque acting on planet in a radiative disk with $\gamma=1.001$ does not reflect the torque acting on a planet in an isothermal disk, as the disk structure is different \citep{2008A&A...478..245P, 2012A&A...546A..99K}. 
 On the other hand, the torque has strong fluctuations in time, indicating that the planet has reached the convective zone of the disk.

In Fig.~\ref{fig:AdiconstGamma}, the radial torque density $\Gamma(r)$, which is the torque exerted by an annulus of width $dr$ and related to the total torque by
\beq
       \Gamma_{\rm tot} = \int_{r_{\rm min}}^{r_{\rm max}} \, \Gamma(r) \, dr \ ,
\eeq
is displayed for $20 M_{\rm Earth}$ planets embedded in fully radiative disks with a constant adiabatic index.
All displayed torque densities $\Gamma(r)$ (with $\gamma>1.2$) feature a spike just below $r \approx 1.0 a_{\rm Jup}$ on top of the underlying Lindblad torque curve, which is the normal pattern for a fully radiative disk with a high adiabatic index \citep{2009A&A...506..971K}. 
For increasing $\gamma$, the Lindblad torque and the spike in the torque density are reduced, indicating a reduction of the positive corotation torque. 
This reduction in torque density is also represented by a decrease in the total torque (see Fig.~\ref{fig:AdiTorque}). 
For $\gamma=1.1$, no spike in the torque density is visible, and only the Lindblad torque remains. 
This also is reflected in the total torque, which is negative (see Fig.~\ref{fig:AdiTorque}).

\begin{figure}
 \centering
 \includegraphics[width=0.805\linwx]{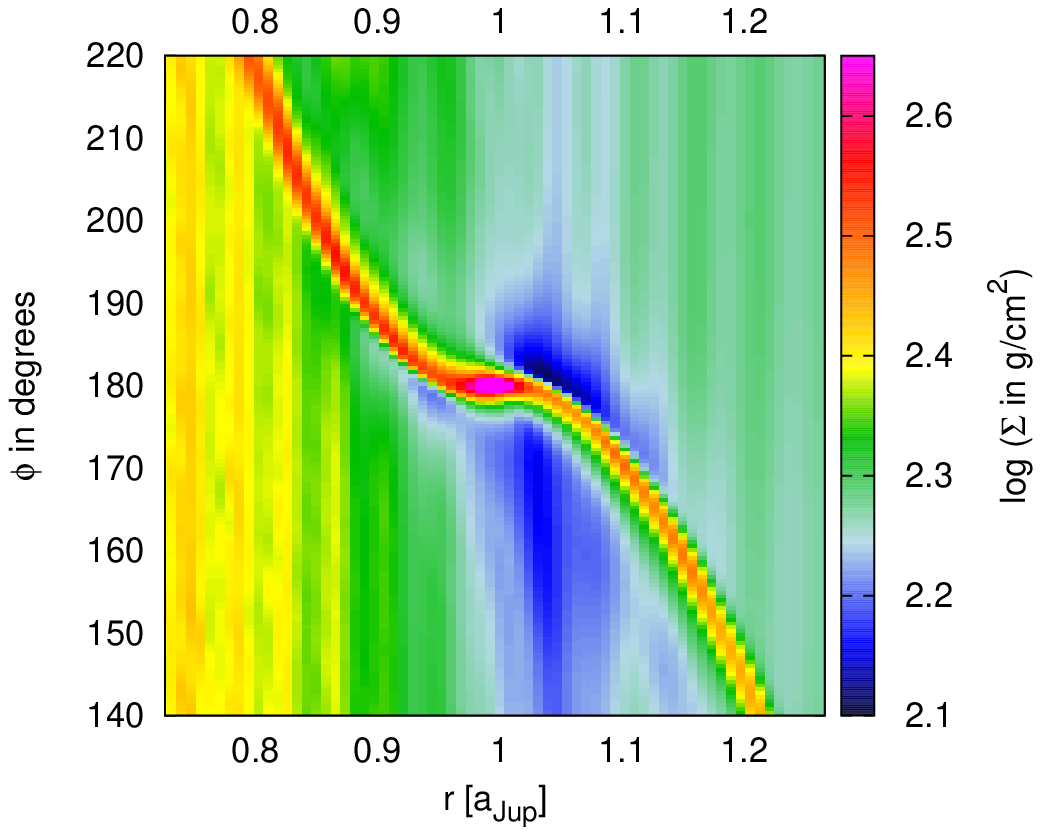}
 \includegraphics[width=0.805\linwx]{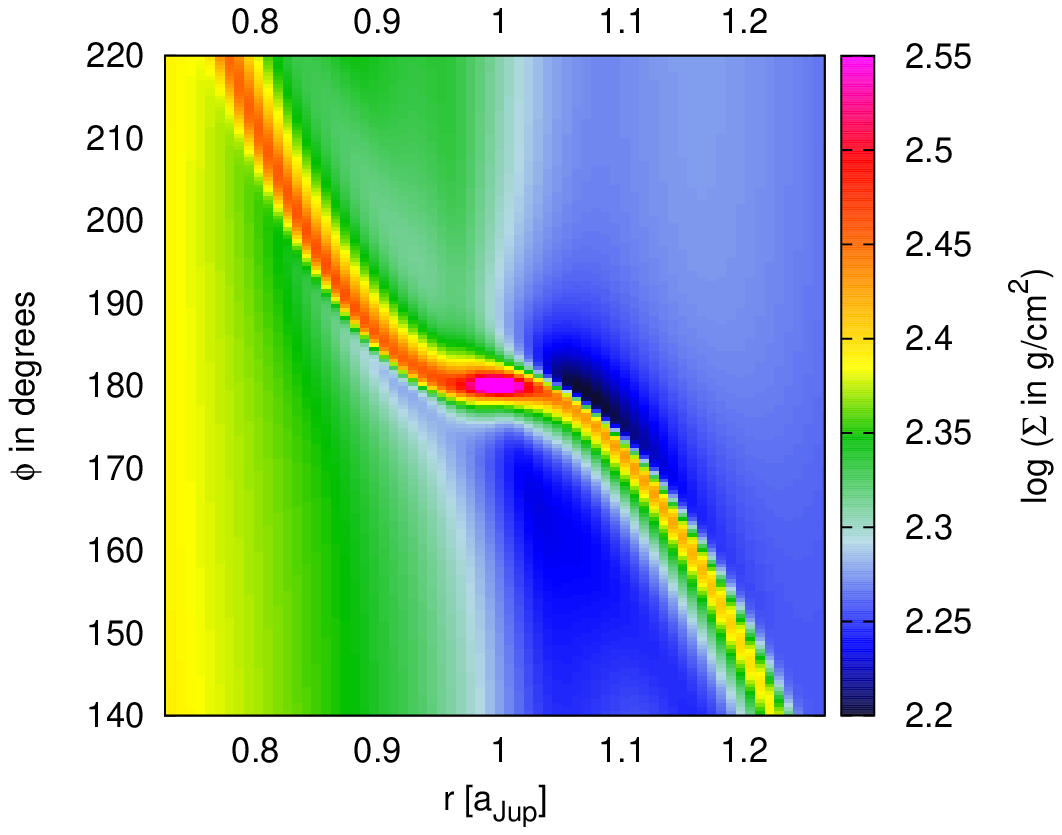}
 \includegraphics[width=0.805\linwx]{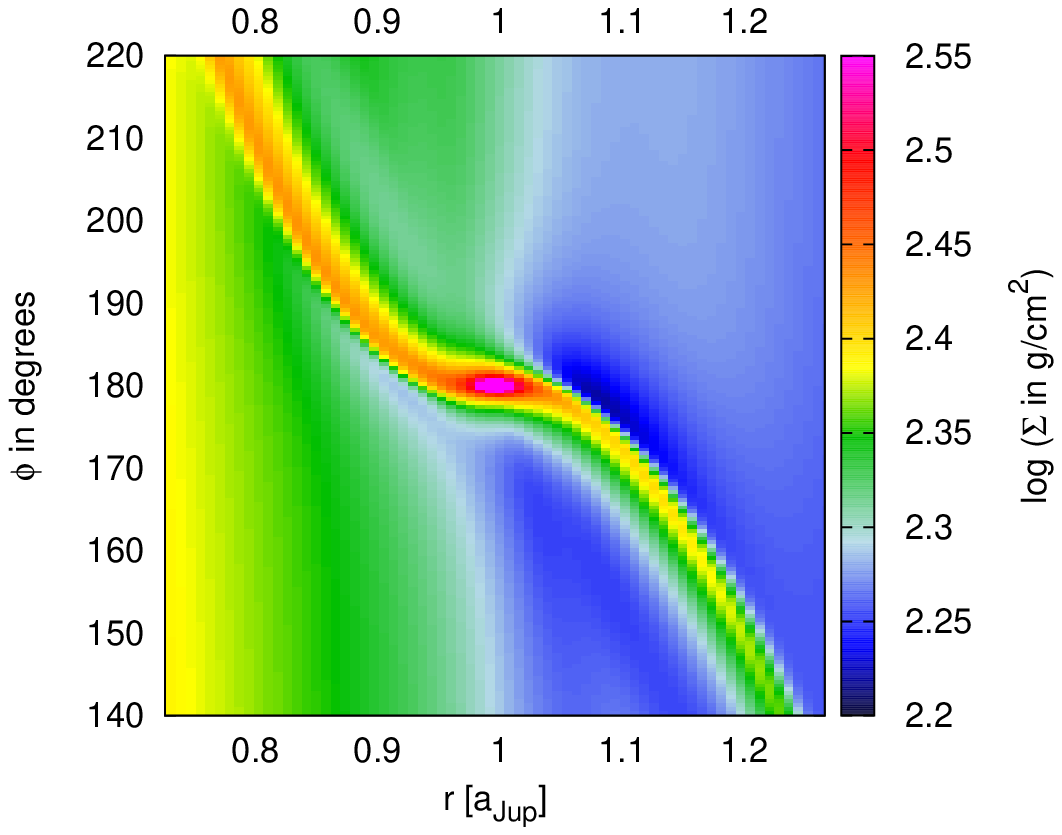}
 \caption{Displayed are the surface density maps for planets on fixed circular orbits in fully radiative disks with different adiabatic indices (top: $\gamma=1.1$, middle: $\gamma=1.4$, bottom: $\gamma=1.7$) and a constant viscosity of $\nu = 10^{15}$\,cm$^2$/s. The disruption in the surface density pattern for the $\gamma=1.1$ disk is due to convection inside the disk.
   \label{fig:Adi2DRho}
   }
\end{figure}

Fig.~\ref{fig:Adi2DRho} shows surface density images of our fully radiative disk simulations with an embedded $20 M_{\rm Earth}$.  
The adiabatic indices for these images are $\gamma=1.1$ (top), $\gamma=1.4$ (middle) and $\gamma=1.7$ (bottom). 
We note that the $\gamma=1.4$ case is discussed in great detail in \citet{2009A&A...506..971K}. 
For the $\gamma=1.1$ simulation, strong variations in the density surrounding the planet are visible. 
The variations are due to convection inside the disk. 

The surface density pattern of the $\gamma=1.7$ disk, on the other hand, shows no sign of convection. In fact it is very similar to the $\gamma=1.4$ (middle in Fig.~\ref{fig:Adi2DRho}). 
The only difference is that the density perturbations in front and behind the planet are not as well pronounced. 
Due to a difference in the adiabatic sound speed, which is $\propto \sqrt{\gamma}$, the opening angles of the spiral waves increase with increasing $\gamma$. 
The density in the spiral waves is reduced for increasing $\gamma$, resulting in a decrease of the Lindblad torque, as can be seen in the torque density distribution (Fig.~\ref{fig:AdiconstGamma}), and in the total torque (Fig.~\ref{fig:AdiTorque}).

\subsection{Influence of viscosity}

\begin{figure}
 \centering
 \resizebox{\hsize}{!}{\includegraphics[width=0.9\linwx]{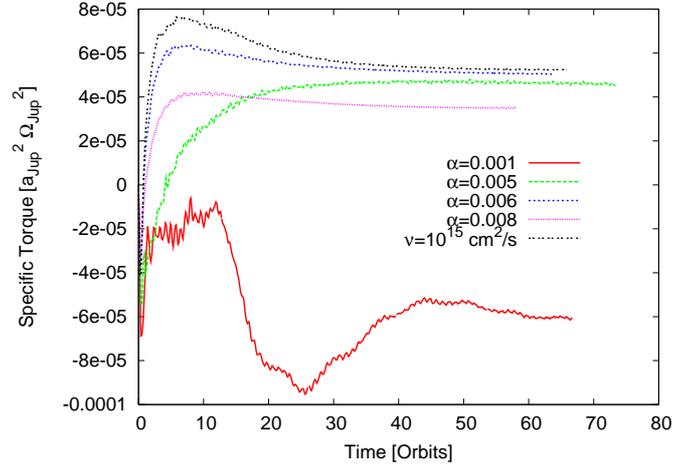}}
 \caption{Torque acting on $20 M_{\rm Earth}$ planets embedded in disks with different viscosities ($\gamma=1.4$).
   \label{fig:alphatorque}
   }
\end{figure}

In Section \ref{subsec:disk-visc}, we discussed the changes of the structure of planetary disks with different viscosities. 
Changes in the temperature gradient lead to changes in the torque acting on embedded planets \citep{2011MNRAS.410..293P}. 
This sensitivity is highlighted in Fig.~\ref{fig:alphatorque}, which shows the total torque acting on an embedded $20 M_{\rm Earth}$ planet in disks with different viscosities. 
Planets in disks with $\alpha \leq 0.002$ (not displayed) feel a negative torque, indicating inward migration. 
A decrease of the torque with decreasing viscosity can also be found in isothermal disks \citep[see for example our simulations in Fig. 1 in][]{2010A&A.523...A30}. 
The decrease in the total torque is a result of the lower viscosity. 
If the viscosity is very low, the corotation torques saturate and outward migration can not be supported any longer  \citep[for isothermal disks, see][]{2001ApJ...558..453M}.
The planet depletes its coorbital region of gas (i.e., begins to open a gap, Fig.~\ref{fig:ARhoall}), which reduces the torque.

For $\alpha$ viscosities with $\alpha \geq 0.003$, the torque acting on the planet becomes positive, and
for even larger viscosities ($\alpha \geq 0.004$), the magnitude of the torque begins to converge. 
The torque acting on the planet with constant viscosity of $\nu = 10^{15}$\,cm$^2$/s is similar to the one with $\alpha \approx 0.006$. 
This is consistent with the similarities between their temperature and density profiles (Fig.~\ref{fig:RhoTemp2D}).


\begin{figure}
 \centering
 \resizebox{\hsize}{!}{\includegraphics[width=0.9\linwx]{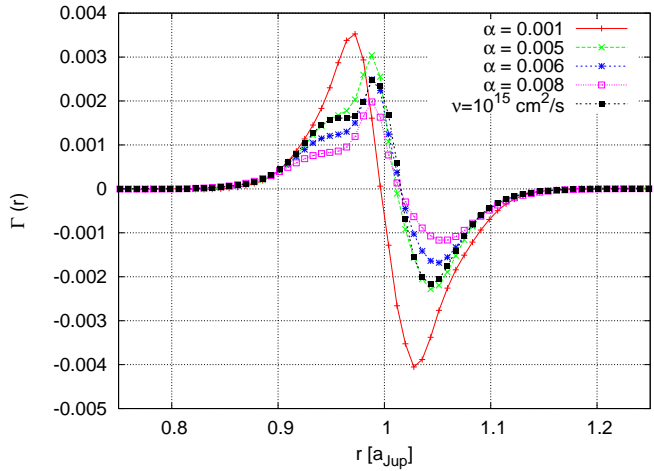}}
 \caption{Profiles of the radial torque density $\Gamma (r)$ acting on $20 M_{\rm Earth}$ planets embedded in disks with different viscosities (in the disk equilibrium state) and a constant adiabatic index ($\gamma=1.4$). 
   \label{fig:alphaGamma}
   }
\end{figure}

\begin{figure}
 \centering
 \includegraphics[width=0.805\linwx]{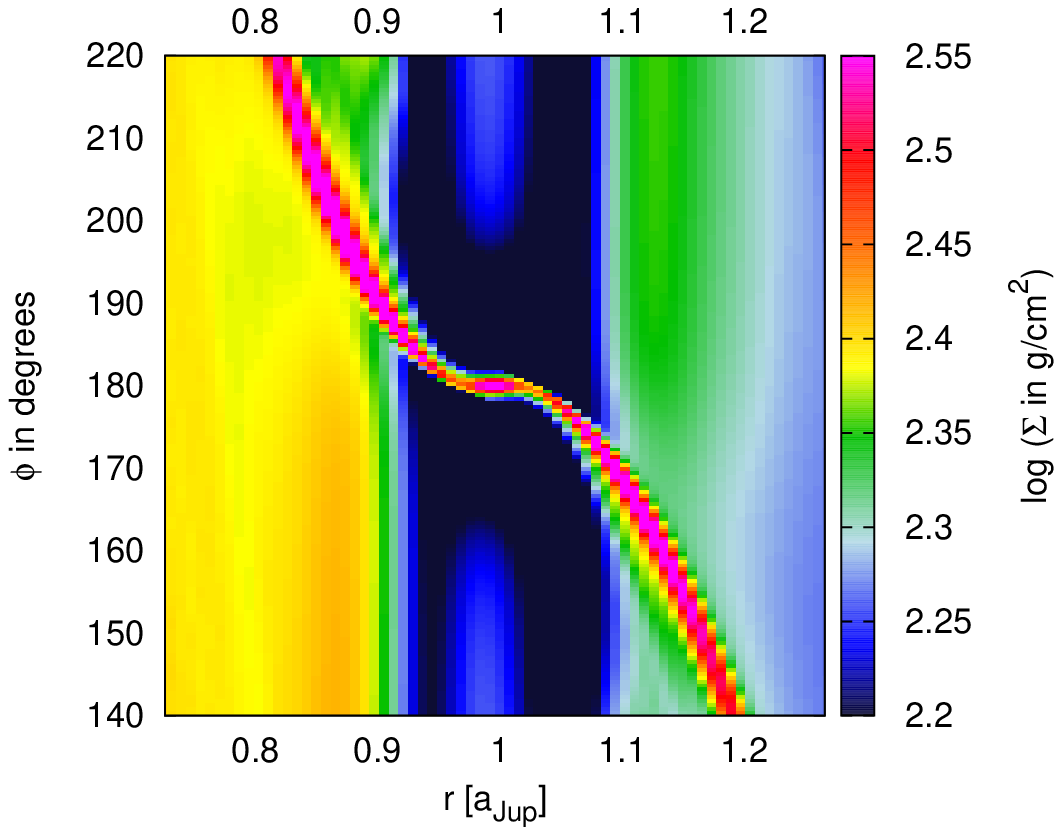}
 \includegraphics[width=0.805\linwx]{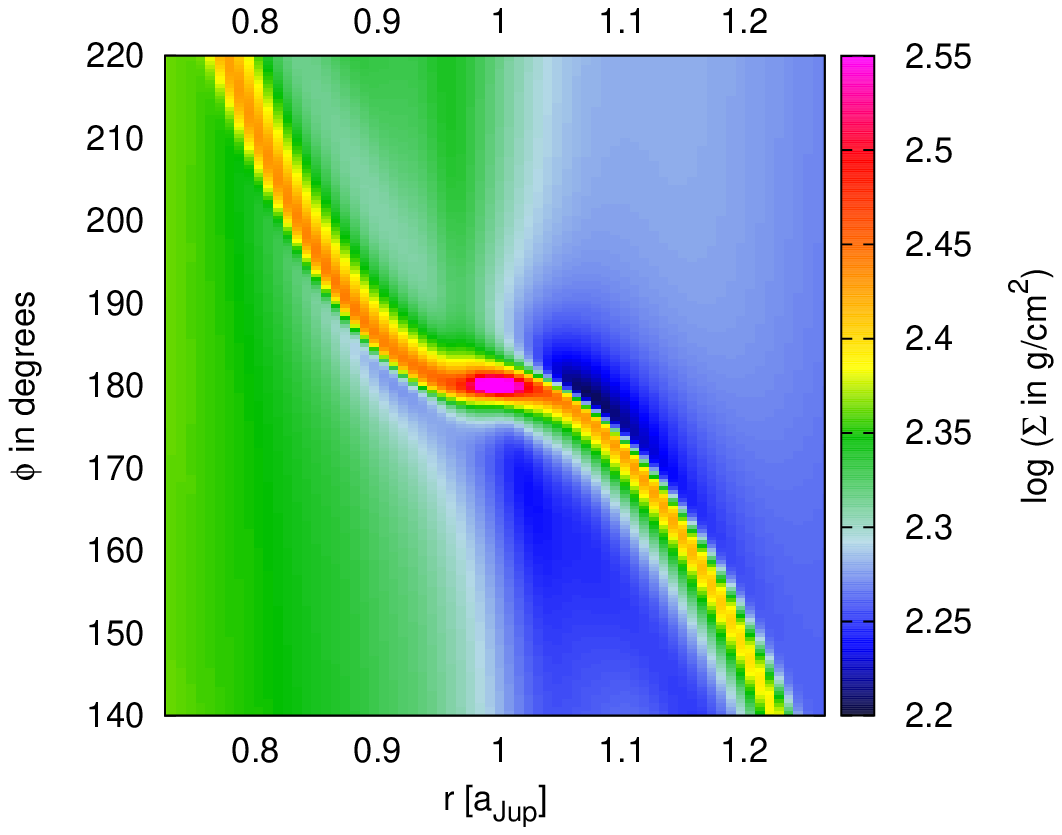}
 \includegraphics[width=0.805\linwx]{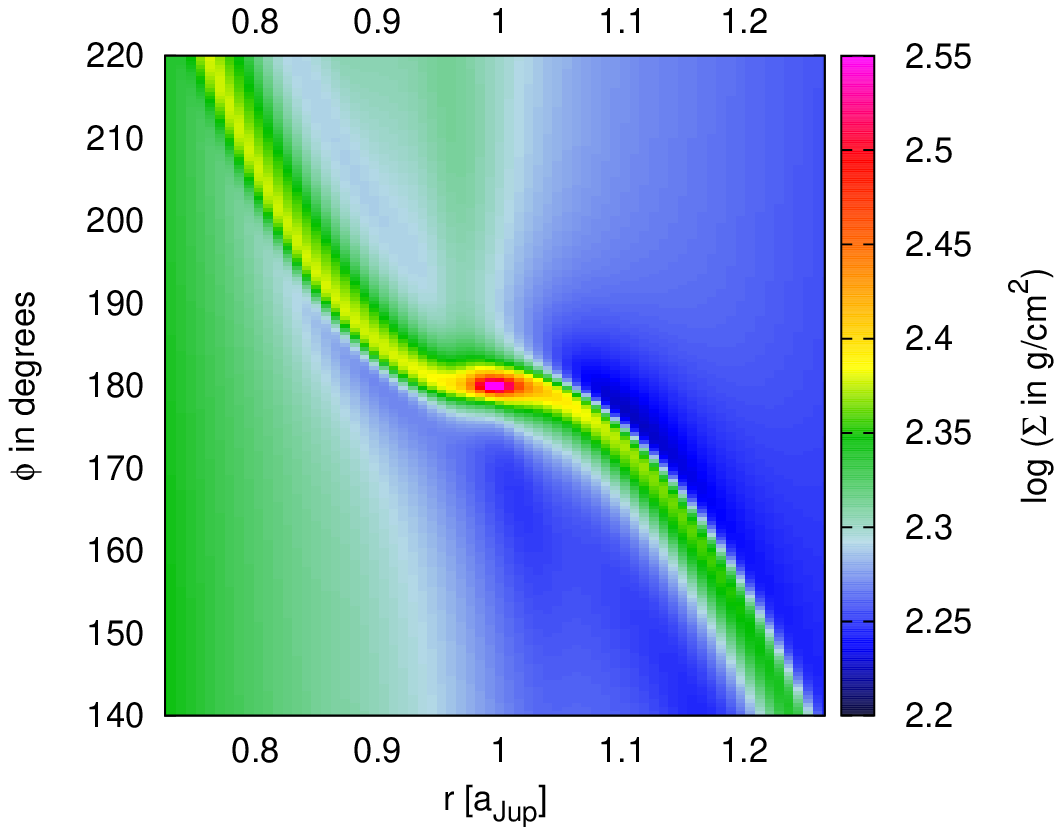}
 \caption{Surface density maps for $20 M_{\rm Earth}$ planets on fixed circular orbits in fully radiative disks (with $\gamma=1.4$). The viscosities in the disks are (from top to bottom): $\alpha=0.001$, $\alpha=0.005$, and $\alpha=0.008$. The snapshots are taken in the equilibrium state.
   \label{fig:ARhoall}
   }
\end{figure}

In Fig.~\ref{fig:alphaGamma} we display the radial torque density $\Gamma (r)$ acting on a $20 M_{\rm Earth}$ planet in disks with different viscosity. 
In the $\alpha=0.001$ simulation, the displayed torque density shows a typical Lindblad torque curve, but without the spike at corotation (see below); a spike in the torque curve at corotation is the typical pattern of a fully radiative disk with a high viscosity \citep{2009A&A...506..971K}. 
In fact the torque density is very similar to an isothermal disk. 

The $\alpha=0.005$ and $\alpha=0.006$ torque density patterns are consistent with the that of the constant viscosity of $\nu = 10^{15}$\,cm$^2$/s. 
The spike in the distribution near $r \approx 1.0 a_{\rm Jup}$ is nearly identical for the constant viscosity and $\alpha=0.006$ simulations, while the $\alpha=0.005$ spike is a little bit higher. 
However, there are some differences. 
For smaller and larger distances from the central star ($r<1.0 a_{\rm Jup}$ and $r>1.0 a_{\rm Jup}$) the torque density is larger for the constant viscosity case compared to $\alpha=0.006$, although all simulations follow the same trend. 
The underlying Lindblad torque distribution is also smaller for the $\alpha=0.005$ disk, due to a difference in the overall disk structure (see Fig.~\ref{fig:RhoTemp2D}). 
The reduction of $\Gamma(r)$ for larger $\alpha$ is partly a consequence of the increased disk temperature. 
These small differences are the reason why the total torque of the constant viscosity simulations is somewhat larger compared to the $\alpha$ viscosity disks.

When $\alpha=0.008$, the Lindblad torque is much smaller compared to the other disks, as the temperature is higher. 
However, the torque density shows the usual spike, indicating a positive corotation torque, which can lead to outward migration if it over-compensates for the negative Lindblad torque \citep{2009A&A...506..971K}. 
One might suspect from the trend of the simulations that even higher viscosities could destroy this effect of outward migration. 
Moreover, modifications to the horseshoe dynamics for large viscosities could limit or cutoff the torque acting on the planet \citep{2001ApJ...558..453M,2002A&A...387..605M}.

In Fig.~\ref{fig:ARhoall} we display the surface densities of several disks  with an embedded $20 M_{\rm Earth}$ planet.  
Different viscosities are shown to gain more insight into the origin of the torque density profiles. 
For very low viscosities ($\alpha=0.001$), the planet seems to open a very small or partial gap inside the disk. 
In addition, the spiral waves of the planet are more dense compared to the constant viscosity simulation (middle panel in Fig.~\ref{fig:Adi2DRho}), which may be due to the higher density in the midplane of the disk.
 A low viscosity encourages the formation of gaps in the disk, thus explaining the torque density in Fig.~\ref{fig:alphaGamma}.

The $\alpha=0.005$ surface density profile seems very similar to the constant viscosity profile we discussed in great detail in \citet{2009A&A...506..971K}. 
The density increase ahead of the planet ($\phi > 180^\circ$ and $r<1.0 a_{\rm Jup}$) is clearly visible in both cases, thus creating a nearly identical 'spike' in the torque density distribution (Fig.~\ref{fig:alphaGamma}). 
However, the density decrease behind the planet ($\phi < 180^\circ$ and $r>1.0a_{\rm Jup}$) is not so clear in the $\alpha=0.005$ simulation compared to the $\nu = 10^{15}$\,cm$^2$/s simulation (middle panel in Fig.~\ref{fig:Adi2DRho}). 
This directly reflects on the torque density, as it is higher for the constant viscosity simulation compared to the $\alpha=0.005$ simulation at that distance from the central star, thus explaining the higher total torque.

For $\alpha=0.008$, the surface density profile shows the same structure as for the constant viscosity simulation, but the surface density is generally reduced. 
The spiral waves and the vicinity near the planet show smaller surface densities. 
The structure ahead and behind the planet is also not as distinctive as in the constant viscosity (or $\alpha=0.005$) simulation. 
This all leads to a smaller curve in the torque density plot (Fig.~\ref{fig:alphaGamma}) and to a smaller total torque (Fig.~\ref{fig:alphatorque}).

\subsection{Varying adiabatic index}

\begin{figure}
 \centering
 \resizebox{\hsize}{!}{\includegraphics[width=0.9\linwx]{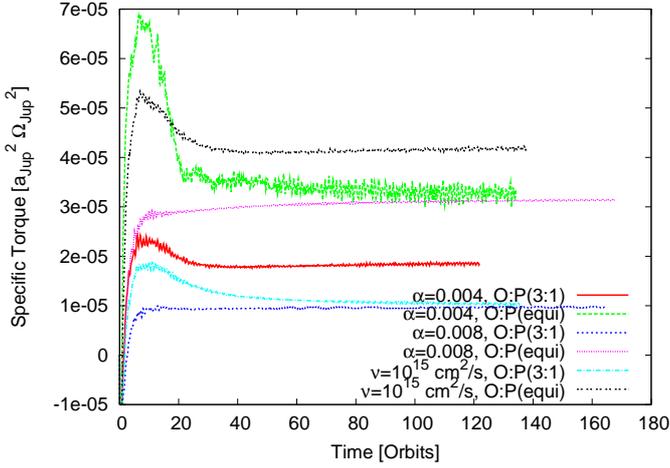}}
 \caption{Torque acting on $20 M_{\rm Earth}$ planets embedded in disks with different viscosities. The graph features $\alpha=0.004$, $\alpha=0.008$, and constant viscosity for both gas ortho-para mixtures. 
   \label{fig:torqueadi}
   }
\end{figure}

In addition to different viscosities, we now consider models that feature a varying adiabatic index.
In Fig.~\ref{fig:torqueadi} the total torque acting on $20 M_{\rm Earth}$ planets in disks is displayed for different viscosity and gas treatments. 
For low viscosities ($\alpha=0.001$) the torque is negative (not displayed), indicating inward migration, and for higher viscosities, the torque is positive indicating outward migration, as expected from simulations with a constant adiabatic index.

For all shown viscosities, the torque acting on the planet is positive, indicating outward migration. 
In the O:P(equi) disk the torque is higher than the torque for a planet in the O:P(3:1) disk (with the same viscosity). 
The largest differences between the two ortho-para gas states is observed for the disks with a constant viscosity. 

The temperature near the planet is in the region of $80 K \leq T_{\rm Planet} \leq 130 K$, which is where the separation of the adiabatic index between O:P(equi) and O:P(3:1) first becomes pronounced for decreasing temperatures (see Fig.~\ref{fig:Adiindex}). 
This difference in temperature leads to a change in the total torque acting on the planet (see Fig.~\ref{fig:AdiTorque}), which can be up to a factor of two in this temperature region. 
Therefore the total torque in the O:P(equi) disk is higher than in the O:P(3:1) disk. 
This effect is due to a change in the adiabatic index in the region near the planet. The origin of this effect is also described in Section~\ref{subsec:CAI}.

\begin{figure}
 \centering
 \resizebox{\hsize}{!}{\includegraphics[width=0.9\linwx]{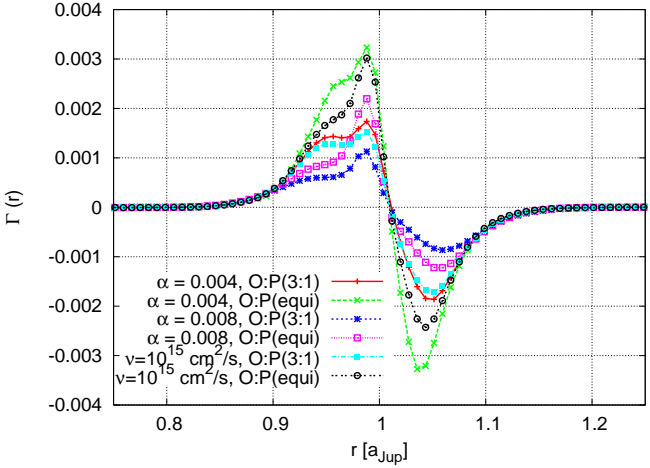}}
 \caption{Torque density of $20 M_{\rm Earth}$ planets embedded in disks with different viscosity. The graph features $\alpha=0.004$, $\alpha=0.008$, and a constant viscosity for both gas ortho-para mixtures.
   \label{fig:Gammaadi}
   }
\end{figure}

Fig.~\ref{fig:Gammaadi} shows the torque density $\Gamma(r)$.
 As all total torques are positive (Fig.~\ref{fig:torqueadi}), each simulation shows a spike on top of the Lindblad torque in the $\Gamma(r)$ distribution just below $r \approx 1a_P$. 
 This spike is largest for O:P(equi) in the constant viscosity simulation, which is also the simulation featuring the highest total torque. 
The spikes in the $\Gamma(r)$ distributions are smaller for O:P(3:1) disks compared to O:P(equi) disks, which is also reflected in the total torque (Fig.~\ref{fig:torqueadi}). 
 However, the Lindblad torque seems strongest for the $\alpha=0.004$ disk with the O:P(equi) gas.

The torque density in the O:P(equi) disk with a constant viscosity is slightly smaller around $r \approx 1.0 a_{\rm Jup}$ compared with the constant viscosity and $\gamma$ case (see Fig.~\ref{fig:alphaGamma}), which leads to a smaller total torque. 
The torque density spike in the disk with O:P(3:1) is much smaller than in the constant $\gamma$ case (both for constant viscosity again), which results in a smaller total torque. 
The same trend of the behavior of the torque density seems to be true for $\alpha$ viscosities with $\alpha=0.008$. 
The spike in the torque density is larger for the O:P(equi) simulations than in the O:P(3:1) simulations. 
This indicates a larger positive corotation torque, which results in a larger torque acting on the planet (see Fig.~\ref{fig:torqueadi}). 
This is a consequence of the temperature dependence of the adiabatic index, where $\gamma$ has smaller values in the O:P(equi) disks, resulting in a larger total torque (see Fig.~\ref{fig:AdiTorque}).

\begin{figure}
 \centering
 \includegraphics[width=0.795\linwx]{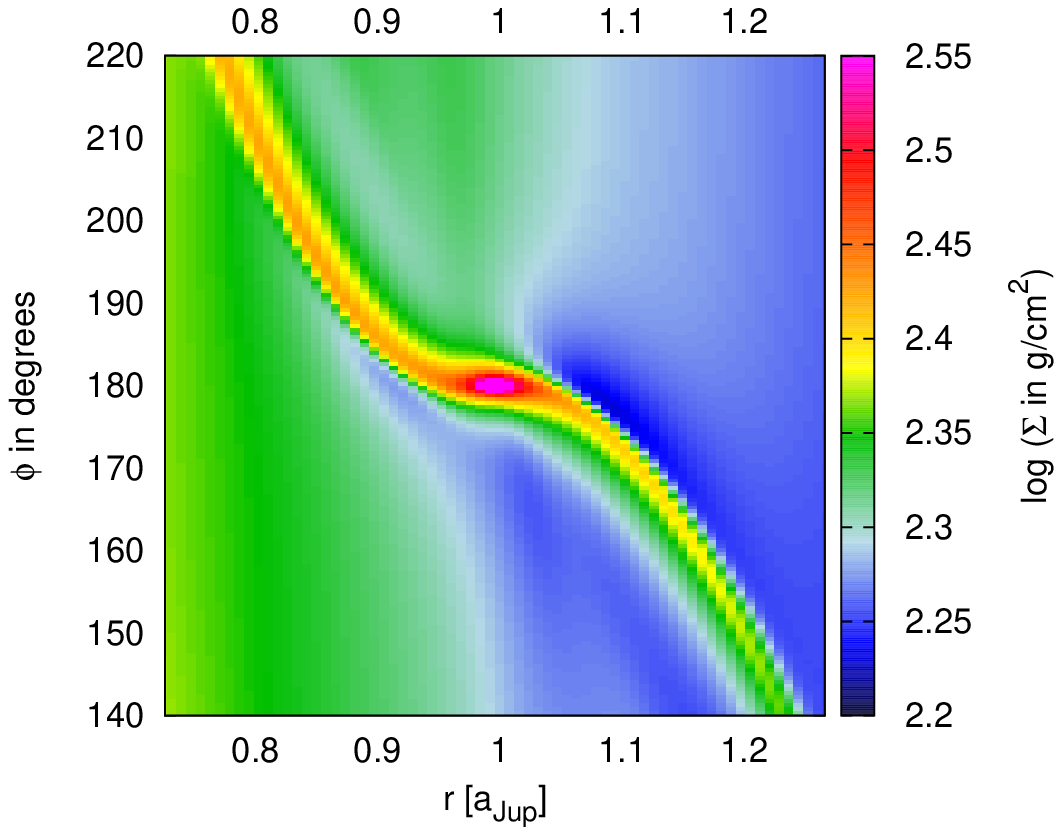}
 \includegraphics[width=0.795\linwx]{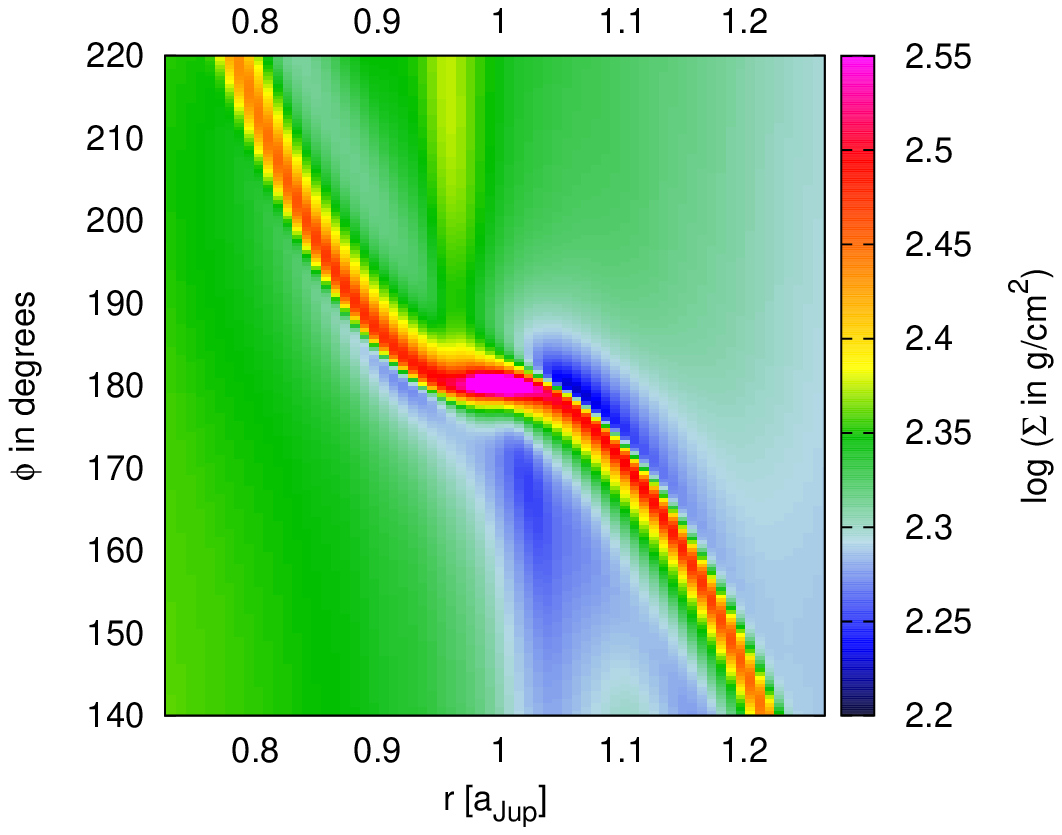}
 \includegraphics[width=0.795\linwx]{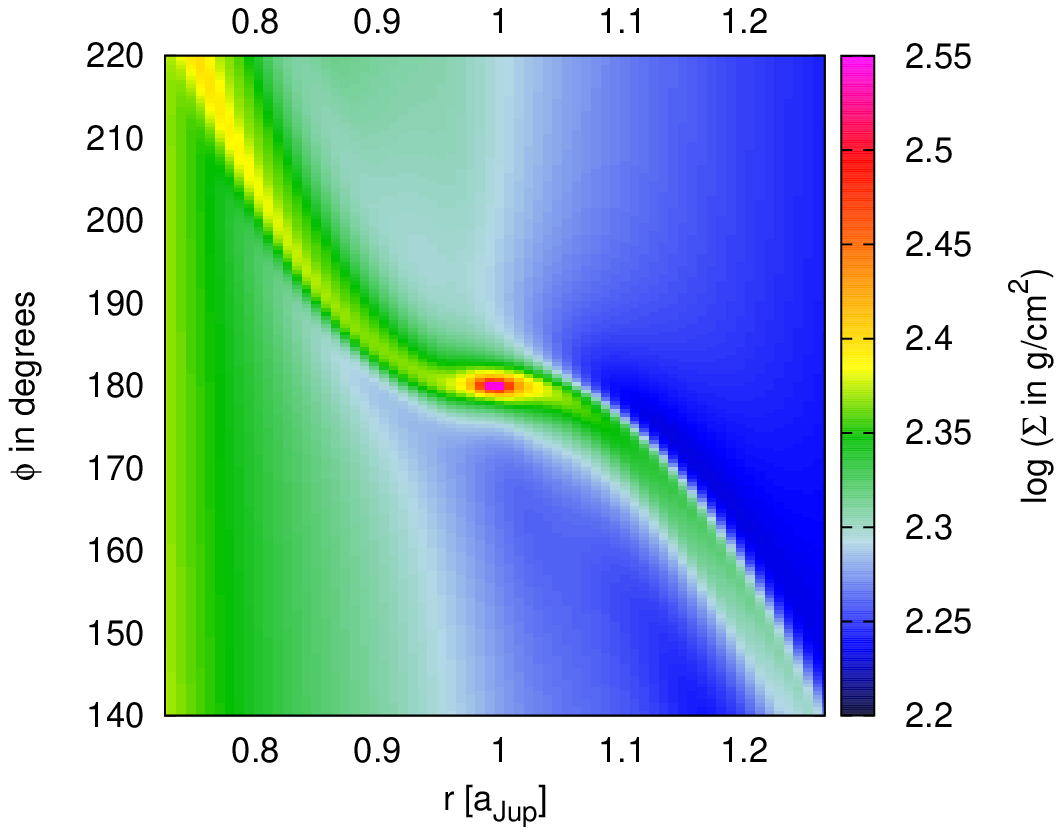}
 \includegraphics[width=0.795\linwx]{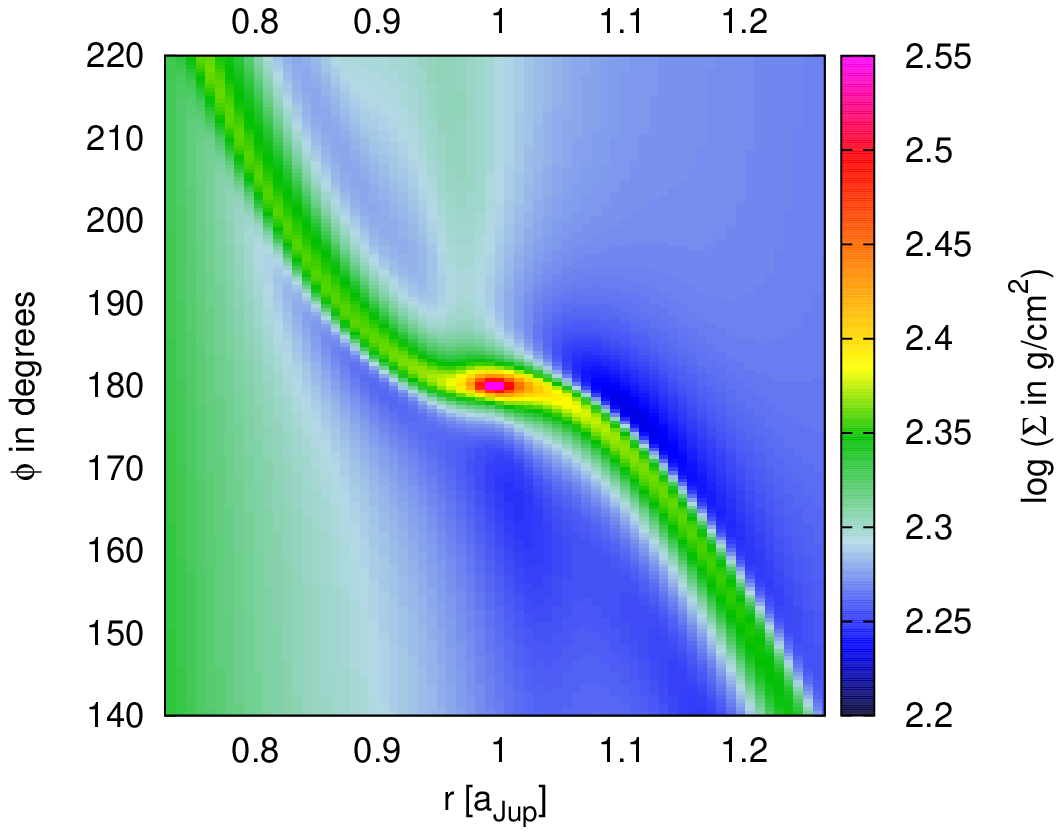}
 \caption{Displayed are the surface density maps for $20 M_{\rm Earth}$ planets on fixed circular orbits in fully radiative disks. The viscosities and gas mixtures in the disks are (from top to bottom): $\alpha=0.004$ with O:P(3:1), $\alpha=0.004$ with O:P(equi), $\alpha=0.008$ with O:P(3:1), and $\alpha=0.008$ with O:P(equi). The snapshots are taken in the heating/cooling equilibrium state of the disk.
   \label{fig:Rho0048adiall}
   }
\end{figure}

In Fig.~\ref{fig:Rho0048adiall} the surface density for the $\alpha=0.004$ and $\alpha=0.008$ simulations with the two different ortho-para gas ratios are displayed. 

The surface density is a little bit higher in front of the planet ($\phi > 180$ and $r<1.0 a_{\rm Jup}$) and a little bit lower behind the planet ($\phi < 180$ and $r>1.0 a_{\rm Jup}$) for O:P(equi) compared with O:P(3:1) for both of the viscosities shown, influencing the positive corotation torque. 
This can also be seen in the $\Gamma(r)$ distribution (Fig.~\ref{fig:Gammaadi}), where the spike becomes larger, resulting in the observed higher total torque for the O:P(equi) models.

For the lower viscosity ($\alpha=0.004$), the density in the spiral waves exerted by the planet is higher compared to the high viscosity case ($\alpha=0.008$), which could also be observed for disks with a constant $\gamma$ (Fig.~\ref{fig:ARhoall}). 
Also, the surface density near the planet is higher for the lower viscosity disks. These two features cause the higher torque in the $\alpha=0.004$ disks compared with $\alpha=0.008$  (for the same gas ortho-para ratios).

When $\alpha=0.004$, the spiral waves in the O:P(equi) disks are much denser compared with O:P(3:1) disks, but it seems to be the other way around for $\alpha=0.008$. The difference in the spiral wave density influences the height of the Lindblad torques in Fig.~\ref{fig:Gammaadi}.





In total, the differences in disk structure between the two different ortho-para mixtures are larger in a constant viscosity disk than in disks with an $\alpha$-viscosity. The reason for this may be in the complicated relation that arises when an $\alpha$ viscosity is used, because in this case $\nu$ depends explicitly on the temperature. 

\section{Zero-torque radius}
\label{sec:zerotorque}

The torque acting on an embedded $20M_{\rm Earth}$ planets decreases with increasing distance from the central star \citep{2011A&A...536A..77B}. 
At some point in the disk, the torque becomes zero (zero-torque radius $R_0$) and the planet ceases to migrate due to planet-disk interactions. 
The zero-torque radius $R_0$ is strongly dependent on planetary mass and on gradients in the disk. In addition the location of $R_0$ depends on the mass of the disk \citep{2011A&A...536A..77B}, and based on the results presented in Section~\ref{sec:influence}, we also expect $R_0$ to depend on the viscosity and the equation of state of the gas. 

\begin{figure}
 \centering
 \resizebox{\hsize}{!}{\includegraphics[width=0.9\linwx]{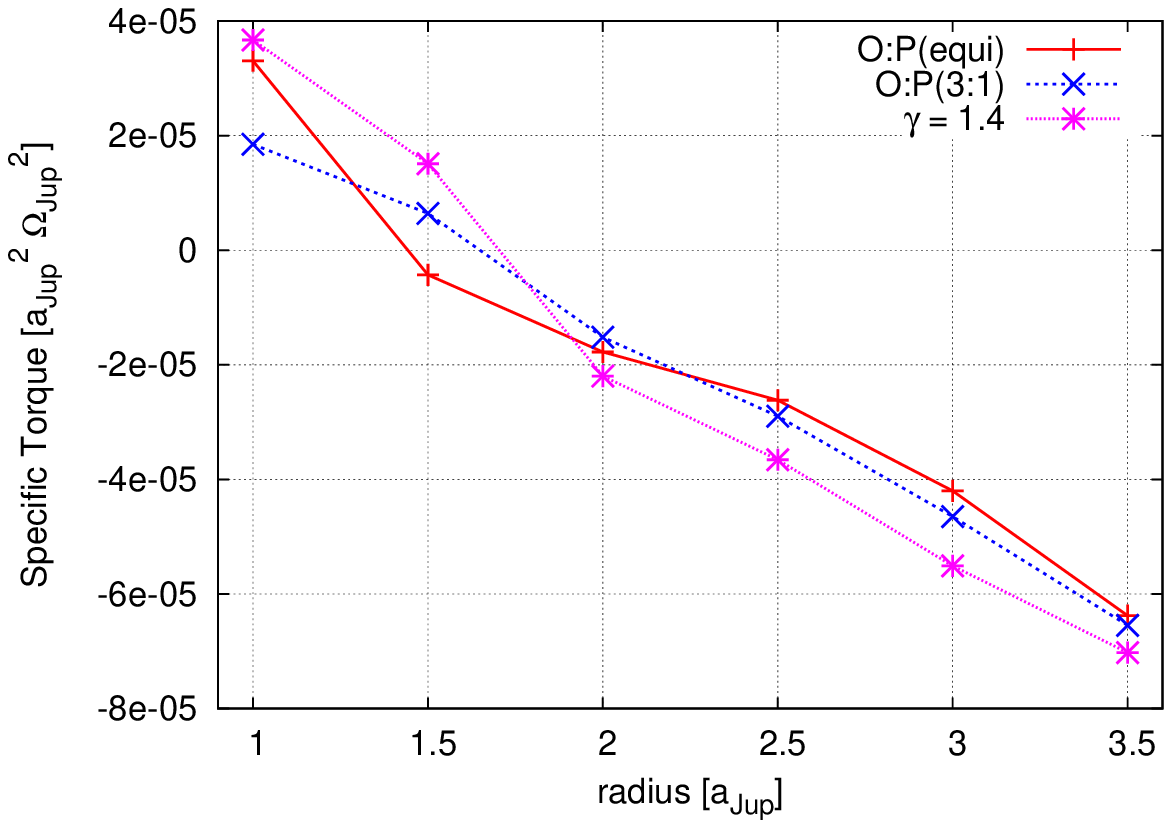}}
 \resizebox{\hsize}{!}{\includegraphics[width=0.9\linwx]{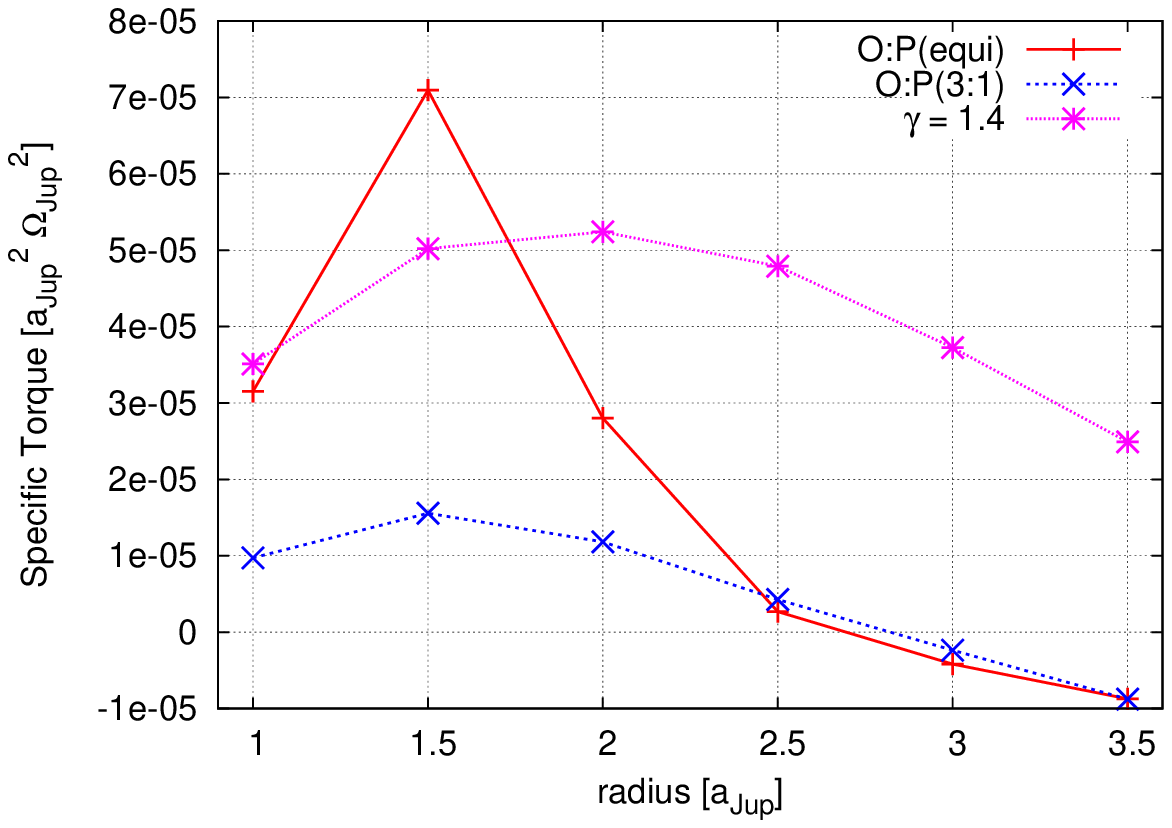}}
 \resizebox{\hsize}{!}{\includegraphics[width=0.9\linwx]{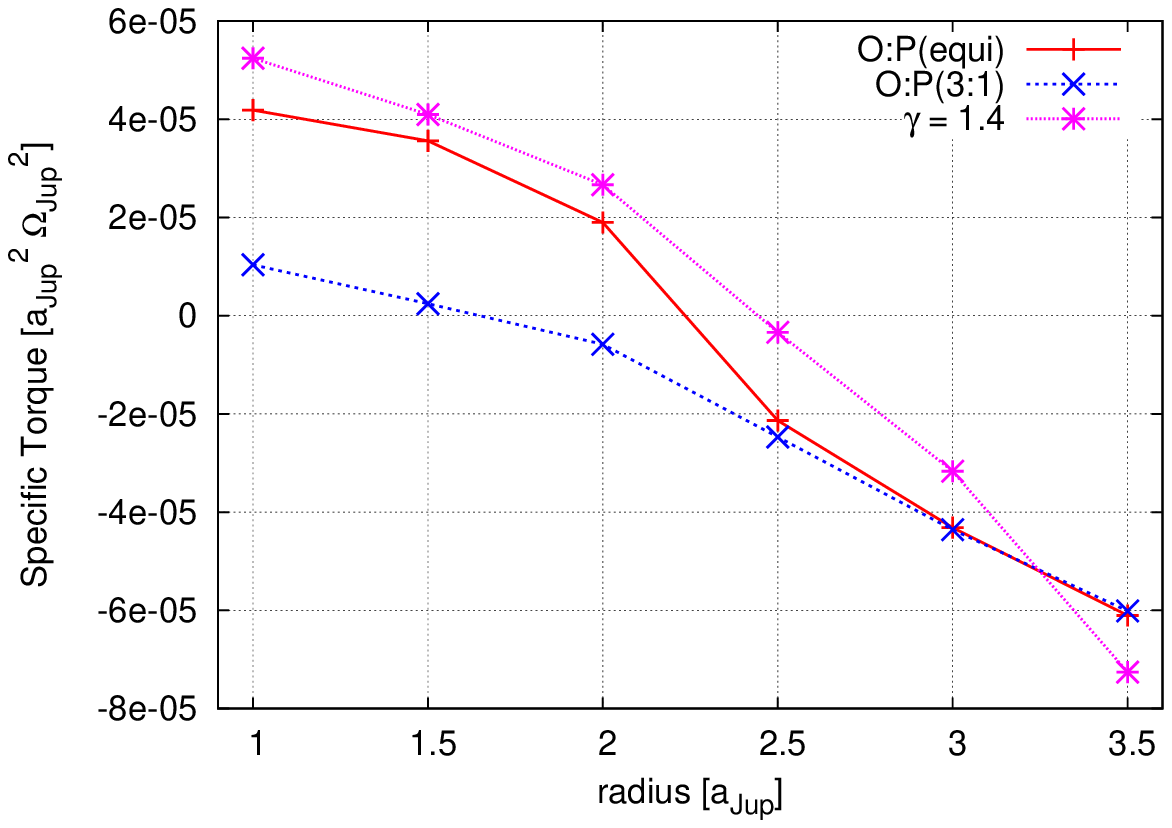}}
 \caption{Torque acting on $20 M_{\rm Earth}$ planets in disks with different viscosities. Each plot features the two gas mixtures [O:P(3:1) and O:P(equi)] and a constant $\gamma=1.4$. From top to bottom the viscosity is: $\alpha=0.004$, $\alpha=0.008$, and $\nu = 10^{15}$\,cm$^2$/s. The simulations with $\gamma=1.4$ and a constant viscosity are taken from \citet{2011A&A...536A..77B}.
   \label{fig:alpnutorque}
   }
\end{figure}

In Figure~\ref{fig:alpnutorque} we show the torque acting on $20 M_{\rm Earth}$ planets placed at various radii in fully radiative disks with different viscosities. 
The figure also features simulations with O:P(equi), O:P(3:1), and a constant adiabatic index $\gamma=1.4$. 
The data of the simulations with constant $\gamma$ and constant viscosity are taken from \citet{2011A&A...536A..77B}. 
Because a high viscosity is needed to keep torques that are acting on a planet from saturating, high-viscosity disks should have their $R_0$ at larger orbital distances than low-viscosity disks. 
This is indeed the case in Figure~\ref{fig:alpnutorque}, where $R_0$ for the $\alpha=0.004$ simulation is closer to the star than in either the $\alpha=0.008$ or $\nu = 10^{15}$\,cm$^2$/s simulation. 
For these viscosities, $R_0$ is usually largest for the constant adiabatic index simulations and is smallest for the O:P(3:1) simulations. 
The most realistic treatment for the equation of state is likely O:P(3:1), so the other approximations tend to overestimate the location of $R_0$. Table~\ref{tab:zerotorque} summarizes our findings.

To understand these results in more detail, consider Figures~\ref{fig:Adiindex} and \ref{fig:AdiconstGamma}. 
The torque acting on an embedded $20 M_{\rm Earth}$ planet decreases with increasing adiabatic index. 
At low temperatures, the adiabatic index for O:P(3:1) is constantly increasing with increasing temperature over the range of temperatures of interest.
 For O:P(equi), the adiabatic index decreases until $\gamma \approx 1.3$ and then increases until $\gamma=1.67$ at $T \approx 20K$. 
 As the temperature decreases with increasing distance in the disk, the result that O:P(equi) simulations have a larger $R_0$ than in the O:P(3:1) simulations is consistent with results from the previous sections. 
 Using the same reasoning, the constant adiabatic index simulations should have larger $R_0$ than the O:P(3:1) simulations, and usually have larger $R_0$ than in the O:P(equi) case, although the latter is complicated by the break in monotonicity of the O:P(equi) $\gamma$ temperature dependence.

{%
\newcommand{\mc}[3]{\multicolumn{#1}{#2}{#3}}
\begin{table}[b]
 \centering
 \begin{tabular}{l|l|l|l|}\cline{2-4}
  & \textbf{O:P(equi)} & \textbf{O:P(3:1)} & \textbf{const. $\gamma=1.4$}\\\hline
 \mc{1}{|l|}{\textbf{$\alpha=0.004$}} & 1.4 & 1.7 & 1.75 \\\hline
 \mc{1}{|l|}{\textbf{$\alpha=0.008$}} & 2.7 & 2.8 & $>$3.5 \\\hline
 \mc{1}{|l|}{\textbf{const. $\nu$}} & 2.25 & 1.7 & 2.45 \\\hline
 \end{tabular}
 \caption{Zero-torque radii (in $a_{\rm Jup}$) for planets in fully radiative disks with different viscosities and different thermodynamic treatments of the gas.
 \label{tab:zerotorque}
 }
\end{table}
}%

\section{Summary and Conclusions}
\label{sec:summary}

In this paper we have studied the disk structure due to different viscosities (constant $\alpha$ and constant $\nu$, respectively) and changing adiabatic indices. 
We then investigated how an embedded $20 M_{\rm Earth}$ planet migrates in such disks.

The viscosity determines the structure of the disk. 
We find that the midplane density (temperature) decreases (increases) for increasing $\alpha$-viscosity. 
The aspect ratio of the disk and the slope of the temperature profile $\beta_{\rm T}$ also increase for increasing viscosity. 
The simulation with a constant viscosity of $\nu = 10^{15}$\,cm$^2$/s seems to lie between the $\alpha=0.005$ and $\alpha=0.006$ simulations, which has allowed us to make comparisons with the earlier simulations of \citet{2009A&A...506..971K} and \citet{2010A&A.523...A30}.  

The aspect ratio, $H/r$, of the disk is approximately independent of the index $\gamma$. 
Only for very small values, $\gamma \lsim 1.2$, $H/r$ drops due to the onset of convection in the disk. At very low viscosities, the aspect ratio profile is roughly constant with small fluctuations. However, for higher viscosities, the aspect ratio profile has a complex shape, with a slight increase for increasing $r$ at small disk radii, followed by decreasing ratio at large disk radii. 
A decreasing $H/r$ profile with increasing distance is the result of the temperature dependence of the opacity. 

A change in the aspect ratio influences the migration of embedded planets in radiatively cooling disks. 
Although we find that the direction of migration can be inward or outward, the effect of the $H/r$ profile is broadly consistent with results from isothermal disks, where embedded planets migrate inward faster for smaller $H/r$ \citep{2002ApJ...565.1257T}. 
The slope of the temperature $\beta_{\rm T}$ also influences the torque acting on an embedded planet. In many theoretical formulae \citep{2010ApJ...723.1393M, 2011MNRAS.410..293P} the parameter $\beta_{\rm T}$ is used to calculate the Lindblad torque and the entropy-related corotation torque. 
A change in these parameters therefore changes the torque acting on an embedded planet. 
We note that in all cases, isothermal or radiative, viscosity is required to prevent corotation torques from saturating \citep{2001ApJ...558..453M}. In fully radiative disks, radiative transport/cooling acts to prevent saturation of the entropy related torque \citep{2009A&A...506..971K}.

For $\alpha<0.0025$, we find that the torque acting on a $20 M_{\rm Earth}$ planet is negative (inward migration), consistent with planets embedded in an isothermal disk \citep{2010A&A.523...A30}. 
In the low viscosity case, a low-mass planet seems to open a partial gap in the disk, such that corotation torques are weakened. 
For increasing viscosities ($\alpha>0.003$), the torque becomes positive and seems to stall for $\alpha\gtrsim0.005$. Although the density, temperature, and aspect ratio profiles for the $\alpha=0.005$ simulation match quite well with the constant viscosity simulation of an unperturbed disk, there is a difference of about $50\%$ in the total torque acting on an embedded $20 M_{\rm Earth}$ planet, emphasizing further that the treatment for viscosity affects planet migration. 
Because the $\alpha$-viscosity is dependent on the temperature in the disk ($c_s \propto \sqrt T$), and because the presence of the planet modifies the temperature in its neighborhood, the torques will change. 
Overall, this results in a smaller torque acting on the embedded planet.

In addition to viscosity, the adiabatic index influences the torque on the planet. 
In our test simulations with a constant adiabatic index, the torque on a planet is about a factor of three larger for an equation of state with a fixed $\gamma=7/5$ than for a fixed $\gamma=5/3$. 
For the simulations in which the adiabatic index is allowed to vary with temperature, the torque acting on embedded planets in a disk with O:P(equi) is larger than in simulations with O:P(3:1) for the same viscosity. 
This picture is complicated though by the connection between the viscosity and the adiabatic index. Higher viscosity results in higher midplane temperatures, which affects the temperature-dependent adiabatic index. 
The adiabatic index in turn alters the structure of the disk by changing for example the $H/r$ profile. As a result, the location of the zero-torque radius is moved in the disk.  

If viscosity and disk temperature is positively correlated with disk mass, we would then expect massive disks to support outward migration of embedded planets. 
This may be a way of transporting cores to the outer nebula and maintaining low eccentricities, which may be necessary to explain systems such as Fomalhaut \citep{2008Sci...322.1345K, 2012ApJ...750L..21B} and HR8799 \citep{2008Sci...322.1348M, 2010Natur.468.1080M}. 
However, the embedded planet would need to be prevented from undergoing rapid gas accretion during the migration. 
We suggest that this might be an opacity effect that is exacerbated by the high disk temperatures around A stars. 
The higher temperature will place the zero-torque radius at larger semi-major axes, and the naturally higher mass of A star disks will increase the dust opacity, for constant metallicity. 
This could prevent cooling and rapid gas capture onto the planetary core until much lower surface densities are reached.

In addition to affecting massive planetary cores, the zero-torque radius, could in principle serve as a location for collecting embryos, perhaps aiding in the formation of a large core \citep{2011ApJ...728L...9S}. 
However, variations in viscosity as the disk evolves will cause the location of the zero-torque radius to also evolve. 
This could result in the loss of embryos and changes that result to inward migration, or could open the possibility of providing new feeding zones for young planets. 
The overall influence of the zero-torque radius on the formation of large cores is further complicated by collisions between planetesimals in the envisaged planet trap, which may influence the opacity. 
Such a change will affect the $H/r$ profile, which ultimately influences the zero-torque radius.

The O:P(3:1) equation of state is the most realistic for disks explored here. 
This treatment for the equation of state also leads to the smallest zero-torque radii compared with both the constant $\gamma=1.4$ and the O:P(equi) simulations. 
The effect is most pronounced in high-viscosity disks, and the inclusion of a proper equation of state is crucial for determining the location of possible planet traps, where planetary embryos or cores can collect.

\begin{acknowledgements}

B. Bitsch has been sponsored through the German D-grid initiative and through the Helmholtz Alliance {\it Planetary Evolution and Life}. W. Kley acknowledges the support through the German Research Foundation (DFG) through grant KL 650/11 within the Collaborative Research Group FOR 759: {\it The formation of Planets: The Critical First Growth Phase}. 
The calculations were performed on systems of the Computer centre of the University of T\"ubingen (ZDV) and systems operated by the ZDV on behalf of bwGRiD, the grid of the Baden  W\"urttemberg state. 
A. C. Boley's support was provided by a contract with the California Institute of Technology (Caltech) funded by NASA through the Sagan Fellowship Program.
 We would also like to thank an anonymous referee for the useful comments that helped to improve the manuscript.

\end{acknowledgements}

\bibliographystyle{aa}
\bibliography{kley8}
\end{document}